\documentclass[aps, rmp, amsmath, amssymb, reprint, twocolumn, longbibliography]{revtex4-2}

\usepackage{graphicx}
\usepackage{hyperref}
\hypersetup{hidelinks}
\usepackage{bm}
\usepackage{physics}
\usepackage[dvipsnames]{xcolor}

\newcommand{\rev}[1]{#1}
\newcommand{\thor}{$^{229}$Th}
\newcommand{\thorg}{$^{229g}$Th}
\newcommand{\thorm}{$^{229m}$Th}
\newcommand{\regthor}{$^{232}$Th}
\newcommand{\lisaf}{LiSrAlF$_6$}

\newcommand{\caf}{CaF$_2$}

\def\thf{\thor F$_4$ }

\newcommand{\thorox}{ThO$_2$}

\def\Thppp{\ensuremath{\textrm{Th}^{3+}}}
\def\Thpppp{\ensuremath{\textrm{Th}^{4+}}}

\newcommand{\thcaf}{\ensuremath{^{229}\textrm{Th:CaF}_2}}
\newcommand{\thlisaf}{\ensuremath{^{229}\textrm{Th:LiSrAlF}_6}}

\usepackage{upgreek}

\usepackage[version=3]{mhchem}

\usepackage{tikz}
\usetikzlibrary{calc,arrows,arrows.meta,decorations.pathmorphing,intersections,positioning}
\tikzset{
    font = \sffamily,
    level/.style = {thick,black},
    all_arrows/.style = {thick,->,>=Triangle},
    uv_photon/.style={all_arrows,violet},
    vis_photon/.style={all_arrows,red},
    vuv_photon/.style={all_arrows,blue},
    iso_photon/.style={all_arrows,byzantine}
}

\usepackage{tikz-3dplot}
\usepackage[europeanresistors,americanvoltages]{circuitikz}

\begin{document}

\title{Colloquium: Nuclear clocks}

\author{Andrei Derevianko }
\affiliation{Department of Physics, University of Nevada, Reno, Nevada 89557, USA}
\author{R. Elwell}
\affiliation{Department of Physics and Astronomy, University of California, Los Angeles, CA 90095, USA}
\author{Eric R. Hudson}
 \affiliation{Department of Physics and Astronomy, University of California, Los Angeles, CA 90095, USA}
 \affiliation{Challenge Institute for Quantum Computation, University of California Los Angeles, Los Angeles, CA, USA}
 \affiliation{Center for Quantum Science and Engineering, University of California Los Angeles, Los Angeles, CA, USA}

\date{October 23, 2026}

\begin{abstract}

The \thor{} nuclear isomeric state has the lowest energy of all known nuclear excited states, placing it within the reach of current table-top laser technology. This extraordinary property has made this nuclear isomer an attractive candidate for a nuclear optical clock of incredibly high precision and accuracy, both as isolated trapped \thor{} ions and embedded into solid-state platforms. Activity around \thor{} has surged in recent years, driven by breakthroughs in its direct laser excitation. The underlying nuclear physics that gives rise to this unique isomer will be elucidated, as well as the nearly half-century of efforts that led to its direct excitation. The design and systematics of a \thor{} nuclear clock will be discussed, both in ion traps and in the solid-state. These systematics, such as frequency shifts and quenching channels, can be leveraged both to probe the local chemical environment, and as a control knob during clock operation. Finally, the nuclear clock's high sensitivity to the variations of fundamental constants will be discussed.
\end{abstract}

\maketitle
\tableofcontents

\section{Introduction}
This review is motivated by recent breakthrough observations~\cite{Tiedau2024-caf2,Elwell2024-lisaf,Zhang2024-Th229Comb,Zhang2024-ThF4,Elwell2025-ThO2} of laser excitation of the uniquely low-energy $\sim 8.4 \,\mathrm{eV}$ nuclear isomer transition in \thor{}, enabling coherent manipulation of {\em nuclear} degrees of freedom.
These observations pave the way for realizing a novel class of timekeeping devices, nuclear clocks.
Beyond timekeeping, nuclear clocks are expected to be exquisitely sensitive to a variety of exotic physics and can serve as quantum sensors of the ambient environment, including temperature and local material fields.

\rev{For complementary reviews of the \thor{} isomer and the development of nuclear clocks, see \citet{VonderWense2020,Beeks2021Review,Thirolf2024Review}.
Applications to tests of fundamental physics are reviewed by \citet{Peik2021Review}.}

The field is advancing rapidly and broadening in scope.
It brings together concepts and techniques from materials science, quantum chemistry, condensed-matter physics, nuclear physics, and atomic and optical physics.
Although it shares important conceptual connections with M\"ossbauer spectroscopy, it presents a distinct challenge: the nuclear and electronic energy scales can be comparable, making nuclear--environment interactions
central rather than perturbative.
This convergence of traditionally distinct subfields creates both significant challenges and new opportunities, and it calls for a deeper understanding of how the local electronic and material environment modifies nuclear observables.

One may distinguish between two major nuclear clock platforms~\cite{PieTam03}: (i) trapped Th ions, e.g., \Thppp{}~\cite{CamRadKuz12,Yudin2025a,ZahMatHume2025-ThDM,Yamaguchi2024}  and (ii) \thor-containing solids~\rev{\cite{Rellergert2010,Kazakov2012-performance-229Th-clock,VonderWense2020}}, such as \thor:\caf{}, \thor:\lisaf doped bulk crystals~\cite{Tiedau2024-caf2,Elwell2024-lisaf,Zhang2024-Th229Comb}, or \thf{} and \thorox{} films~\cite{Zhang2024-ThF4,Elwell2025-ThO2}.
The two platforms offer distinct trade-offs between the nuclear clock accuracy and stability.
Solid‐state  clocks offer superior stability by interrogating macroscopic ensembles of nuclei simultaneously and provide a pathway to truly portable, high‐performance timekeeping.
However, solid‐state platforms are prone to inhomogeneous broadening and environment‐dependent shifts, while
trapped‐ion platforms provide superior accuracy but require long integration times to reach their ultimate resolution.

Our recommended properties of the \thor{} nuclear transition are compiled in Tab.~\ref{Tab:Th229m-props}.
The reader must be made aware that most of these properties, such as the transition frequency and radiative lifetime, depend on the material hosting \thor{}.

\begin{table*}[ht!]
\caption{Properties of the \thor{} nuclear-clock transition.
The transition energy of the bare nucleus differs from the clock
frequency measured in an electronic or solid-state environment because
of the isomer shift. The \thcaf{} frequency is the quadrupole-averaged transition frequency measured by VUV-comb spectroscopy.
Values of quadrupole moments are spectroscopic ones. Bare-nucleus  reduced transition strength $B(M1)$ and radiative lifetime  are derived from the weighted average of measured radiative lifetimes in solid-state experiments~\cite{Tiedau2024-caf2,Elwell2024-lisaf}.}
\label{Tab:Th229m-props}
\begin{ruledtabular}
\begin{tabular}{lc}
Quantity & Value  \\
\hline
Atomic and mass numbers & $Z=90$, $A=229$ \\
Ground-state quantum numbers & $5/2^+[633]$ \\
Isomeric-state quantum numbers & $3/2^+[631]$ \\
Bare-nucleus transition energy~\cite{Perera2025PRL} & $8.272(22)\,\mathrm{eV}$ \\
Transition frequency in \thcaf{}~\cite{Zhang2024-Th229Comb} &
$2\,020\,407\,384.335(2)\,\mathrm{MHz}$ \\
Transition multipolarity & Predominantly M1 \\
Bare-nucleus radiative lifetime~\cite{Morgan2025_internal_conversion}  & $2.3(3) \times10^3\,\mathrm{s}$ \\
Reduced transition strength $B(M1)$~\cite{Morgan2025_internal_conversion} & $ 0.042(11)\,\mu_N^2$ \\
Reduced transition strength $B(E2)$~\cite{Bilous2018} & $2.3\times10^3\,e^2\mathrm{fm}^4$ \\
Ground-state rms charge radius~\cite{AngeliMarinova2013}  & $5.756(14)\,\mathrm{fm}$ \\
Difference in mean-square charge radii $\delta\langle r^2\rangle_{m,g}$~\cite{Yamaguchi2024} &
$0.0097(26)\,\mathrm{fm}^2$ \\
Ground-state magnetic moment $\mu_g$~\cite{PorSafKoz2021-Th3plusHFS} & $0.366(6)\,\mu_N$ \\
Isomer magnetic moment $\mu_m$~\cite{Yamaguchi2024} & $-0.378(8)\,\mu_N$ \\
Ground-state {spectroscopic} quadrupole moment $Q_g$~\cite{PorSafKoz2021-Th3plusHFS} & $3.11(2)\,e\,\mathrm{b}$ \\
Isomer {spectroscopic} quadrupole moment $Q_m$~\cite{Zhang2024-Th229Comb,PorSafKoz2021-Th3plusHFS} & $1.77(1)\,e\,\mathrm{b}$ \\
Isomer-to-ground quadrupole-moment ratio $Q_m/Q_g$~\cite{Zhang2024-Th229Comb}  & $0.57003(1)$ \\
Ground-state intrinsic hexadecapole moment $Q_{40}$~\cite{Bemis1988} & $3.69(72)\,e\,\mathrm{b}^2$
\end{tabular}
\end{ruledtabular}
\end{table*}

The review follows the historical development of the field while using that history to introduce the central physics questions.
The early nuclear-spectroscopy work explains why the isomer was so difficult to identify; the clock proposals explain why the same low energy state is useful; and the recent laser-excitation experiments motivate the present discussion of ions, crystals, shifts, quenching, and exotic-physics searches.
We allow these threads to run in parallel when doing so is clearer than separating ``history'' from ``technical discussion''.

The bulk of the review is focused on the solid-state platforms, where most of the recent developments have happened.
In particular we will describe the criteria of selecting \thor{} hosts and highlight properties of some of the materials, such as \thcaf{}, \thor:\lisaf{}, \ce{^{229}ThF4}, and \ce{^{229}ThO2}, where the nuclear clock signal was observed.
We will discuss the recent spectroscopic results in solid-state systems ~\cite{Tiedau2024-caf2,Elwell2024-lisaf,Zhang2024-Th229Comb,Zhang2024-ThF4}, including photo-quenching~\cite{Terhune2025-photo-induced,Schaden2024-Th-quenching}.
On the theory side we will describe computations of various properties, such as doping geometries~\cite{Takatori2025-xafs-ThCaF2}, nuclear lifetimes in crystal environments~\cite{Tkalya2018,Tkalya2000-M1inMedium}, internal conversion~\cite{Morgan2025_internal_conversion,TkalyaSi2020-Th229anion,Karpeshin2007} and electron-bridge rates~\cite{Nickerson2021}, isomer shifts~\cite{Perera2025PRL,SafPorKoz2018-ThIsomerShift}, and electric field gradients~\cite{Zhang2024-Th229Comb}.

Many workers have contributed to the study of \thor{} over the last 75 years.
With the recent rapid development of the field and editorial constraints, we regret that the cited references in this review cannot be comprehensive, rather they constitute a representative cross-section of the literature.

\section{Historic developments and what makes \thor{} isomer special}
\label{Sec:History}

Nuclear isomers were first discovered in 1921 when Otto Hahn realized that two radioactive substances that he had extracted from uranium salts, were actually two nuclear states of \ce{^234Pa}~\cite{Hahn1921a,Hahn1921b}.
Such metastable nuclear excited states, with lifetimes ranging from a few ns to $>10^{17}$~yr, typically arise when decay of the excited state by e.g. $\gamma$-radiation requires high-order multi-pole radiation due to angular momentum constraints\rev{~\cite{Weizsaecker1936}} and are typically denoted by placing a $m$ after the mass number -- e.g. the compound Hahn denoted as Z, we now know as \ce{^234^mPa}.

The nucleus is a many-body system of protons and neutrons whose dynamics ultimately descend from QCD.  For medium and heavy nuclei such as \thor{}, a direct calculation from QCD is not yet practical; one instead uses effective descriptions in which nucleons move in a mean field and interact through residual forces~\cite{hammer_konig_review_eft}.
Two limiting pictures are especially useful.
Collective models describe the nucleus as a correlated, deformable object, as in the Bohr--Mottelson liquid-drop model
~\cite{bohr_mottelson}.
By contrast, shell models describe individual protons and neutrons filling quantized orbitals in an average potential, with spin-orbit coupling and residual interactions producing the observed magic numbers and low-lying spectra~\cite{casten_2001}.
The \thor{} clock transition sits precisely where these two pictures meet: it is a single-neutron configuration change embedded in a deformed, collectively rotating nucleus.

\begin{figure}[t!]
    \centering
    \includegraphics[width=1\linewidth]{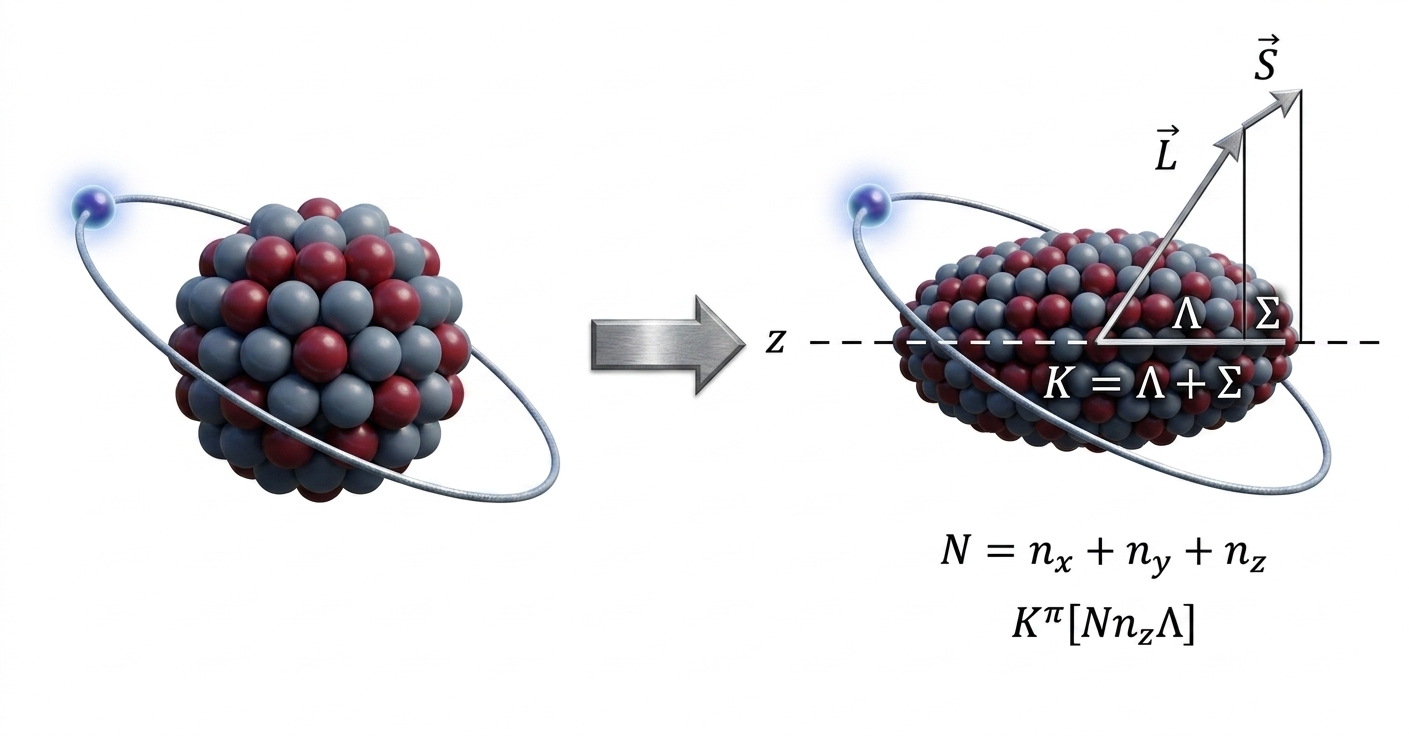}
    \caption{\rev{Schematic progression from a spherically symmetric nucleus (left) to a deformed nucleus (right), introducing the Nilsson labels $K^\pi[N n_z\Lambda]$ used for the \thor{} ground and isomeric configurations.
    The asymptotic projections of the valence neutron's orbital and spin angular momenta, $\vec{L}$ and $\vec{S}$, onto the symmetry axis $z$ are $\Lambda$ and $\Sigma$ (in units of $\hbar$), with $K=\Lambda+\Sigma$.
    Here $\pi$ is parity, $N=n_x+n_y+n_z$ is the total number of oscillator quanta, and $n_z$ counts those along $z$.}}
    \label{fig:nilsson_cartoon}
\end{figure}

{
For atomic physicists, the key idea is close to the electronic shell model, but with two important nuclear twists.  First, the potential is not Coulombic; it is a short-range mean field generated by the other nucleons.
Second, away from magic numbers, valence protons and neutrons can generate strong quadrupole correlations and configuration mixing; when these deformation-driving correlations overcome pairing, the nuclear mean field may become deformed rather than spherical~\cite{casten_2001}.
}
A valence neutron then moves in a deformed potential rather than in a central one.
Nilsson's model captures this situation by starting from a deformed oscillator plus spin-orbit and $\ell^2$ terms~\cite{nilsson_1955} -- here $\ell$ is the orbital angular momentum of the valence neutron.
The resulting single-particle states are labeled by $K^\pi[N n_z\Lambda]$: $K$ is the projection of the total single-particle angular momentum on the symmetry axis, $\pi$ is parity, $N$ is the oscillator shell, $n_z$ counts oscillator quanta along the symmetry axis, and $\Lambda$ is the projection of orbital angular momentum, see Fig~\ref{fig:nilsson_cartoon}.
The deformed core also rotates, so each Nilsson ``bandhead'' is accompanied by a ladder of rotational states.

Within this language the \thor{} isomeric transition is between the excited $3/2^+[631]$ and the ground $5/2^+[633]$ states.
Its energy, now known to be $\sim 8.4$~eV in solid-state environments, is microscopic on the scale of ordinary nuclear excitation energies, which are typically keV to MeV.
The low energy is an accidental near degeneracy of two deformed-neutron configurations, which can be understood in two ways.
First, the change in the contributions of the strong and electromagnetic interactions between the two states nearly cancel, $\Delta E=\Delta E_{\rm strong}+\Delta E_{\rm EM} \approx 8.4$~eV.
That cancellation is the reason that the transition is both laser accessible and possesses a fractional variation that is exceptionally sensitive to changes of the fundamental constants governing the electromagnetic and strong interactions~\cite{He2007,flambaum_quark_mass_2009, safronova_clock_review_2019}.
Second, the near degeneracy is anticipated by the Nilsson model for a nucleus with a quadrupole deformation parameter close to that of \thor{}~\cite{ElwellPhD2024}.
Specifically, the measured spectroscopic quadrupole moments of both the $5/2^+[633]$ ground state and $3/2^+[631]$ excited state (see Table~\ref{Tab:Th229m-props}) correspond in the constant-volume approximation to a Nilsson deformation of order $\delta\sim0.19$~\cite{LOBNER1970495}. Standard parameter values of the Nilsson model within this mass range place these deformations near a crossing of the $5/2^+[633]$ and $3/2^+[631]$ orbitals \rev{(see Fig.~5 of~\citet{nilsson_1955} and Fig.~8 of~\citet{jain1990_rev_mod_deformed_nuclei})}.
{Thus, \thor{} is a deformed odd-neutron nucleus in which two Nilsson bandheads happen to fall within an optical photon energy of each other.
Other nuclear-clock candidates exist in different regimes, notably the 12.4~keV isomer in \ce{^45Sc}, which has recently been driven resonantly with an x-ray free-electron-laser~\cite{Shvydko2023}.
At present, however, no other nucleus is known or expected to share the defining \thor{} feature: an optical-scale near degeneracy of two deformed-nuclear bandheads.}

Given that nearly 2000 isomers with lifetime $> 100$~ns are known~\cite{Thoennessen2026} and the $\alpha$-decay of \ce{^233U} into \ce{^229Th} has been studied since the Manhattan Project, it is natural to wonder why the \ce{^229^mTh} isomer was not excited until very recently.
The answer is one of scale.
\ce{^229^mTh} is unique amongst all known isomers as, by nuclear physics standards, the state is nearly degenerate with the ground state.
This small energy separation meant that standard tools for measuring nuclear energy levels, such as $\gamma$-ray spectroscopy, simply could not resolve the two states.
As a result, the assignment and understanding of the nuclear structure of \ce{^229Th} took several decades to resolve; a history which we briefly summarize here.

{\em The first hints: 1947 - 1976.}  In the decades following the second World War, thanks to advances in radiochemistry and detector technology during the Manhattan Project, there was a flurry of activity in the study of radionuclides.
Early studies~\cite{Studier1947,West1952} identified several $\gamma$-ray emission lines, following \ce{^233U} $\alpha$-decay, between 40~keV - 80~keV, and were attributed to excited states of \ce{^229Th}.
In the following quarter century, a number of experiments were undertaken to understand the nuclear structure of \ce{^229}Th by studying the \rev{$\alpha$-emission~\cite{Goldin1956,Trojan1967, Baranov1967}}, conversion electron~\cite{Tretyakov1960,Andersen1961}, and $\gamma$-emission channels~\cite{Ton1970, Cline1971}.
While these studies largely agreed that the $I^\pi = 5/2^+$ nuclear ground state corresponded to the $(5/2)^+[633]$ band head, there was a lack of consensus in identifying some low-lying excited states and tension with Nilsson model predictions~\cite{Ton1970}.
Based on $\gamma$-ray spectroscopy performed under the PhD thesis work of L.A. Kroger~\cite{Kroger1971}, a solution to this puzzle was posited: a second rotational band, described as $(3/2)^+[631]$, began within $\approx 0.1$~keV of the ground state ~\cite{Kroger1976}!

{\em \rev{Early searches and nuclear-clock proposals: 1977--2006.}}
While this proposal resolved the known issues in \ce{^229Th} level assignment, it was indirect as the energy resolution of the $\gamma$-ray spectrometer employed by Kroger and Reich was around 450~eV and therefore prevented a direct observation of the $(3/2)^+[631]$ band head.
Almost a decade later, using a $\gamma$-ray spectrometer of similar resolution and improved calibration, Reich and co-workers performed more accurate $\gamma$-ray spectrometry~\cite{Reich1984}.
However, the motivation for the work was to aid identification of $^{233}$U and no attempt was made to use the improved data to better constrain the location of the $(3/2)^+[631]$ state.
Six years later Reich and Hemler reanalyzed the \ce{^229Th} energy levels using this data and placed the $(3/2)^+[631]$ state at $-1\pm4$~eV relative to the $(5/2)^+[633]$ ground state~\cite{Reich1990}.
They concluded, however, given that studies of \ce{^229Th} decay had established the decay state as $(5/2)^+[633]$, the $(3/2)^+[631]$ state could only be the ground state if the $\gamma$ transition between it and the $(5/2)^+[633]$ state had a lifetime that greatly exceeded $10^4$~yr - in a later work Helmer interpreted this result as bounding the energy of the $(3/2)^+[631]$ state to less than 10~eV~\cite{Helmer1993}.
They then made an estimate of the radiative lifetime between the states as $\sim 7$~hr, given a 1 eV, magnetic dipole transition -- making the prescient comment that their estimate ignores effects of atomic electrons.
This appears to be the first estimate of the excited state lifetime and first indication of the low-lying isomeric state in \ce{^229Th}.

This work caught the attention of a larger audience and in short order several important works appeared.
A few months later, additional evidence in support of the low energy isomer from \rev{$\ce{^230Th}(\mathrm{d},\mathrm{t})\ce{^229Th}$} reactions was presented~\cite{Burke1990} -- this work also makes, what appears to be, the first mention of optically exciting the nucleus.
Almost immediately, Tkalya proposed methods for exciting the nucleus via an inverse electronic bridge mechanism in a plasma~\cite{Tkalya1990} and by optical means~\cite{Tkalya1992} to better measure the isomer energy.
Strizhov and Tkalya also considered the types of decay channels -- internal conversion, electron bridge, and radiative decay -- that could relax the isomer and estimated their properties in atomic thorium~\cite{Strizhov1991}, noting in a prescient comment ``If one of these channels is actually realized, physicists will in time acquire a high-accuracy tool for the determination of subtle effects in atoms, in the structure of a solid, and others."
Similar proposals, leveraging internal conversion and electron bridge processes, appeared around the same time~\cite{Berger1992, Karpeshin1992}, with Kalman also further considering direct optical excitation of the nucleus~\cite{Kalman1992,Kalman1994}.

Given the growing interest in the isomer, Hemler and Reich repeated and improved their $\gamma$-ray spectroscopy measurement by recording more lines with better statistics, using more calibrations and detectors, and improved fitting~\cite{helmer_reich_1994}.
They determined the isomeric state to lie near $3.5$~eV above the ground state.  Although the uncertainty was still large on the optical scale, this result was decisive historically: it shifted the question from whether an anomalously low state existed to how one could find and excite it directly.

This refined measurement was followed by a refinement of the various theoretical proposals for excitation~\cite{Typel1996, Varlamov1996, Karpeshin1996} and study of the nuclear properties~\cite{Dykhne1996}.
In particular, Tkalya studied the mechanism of optical excitation of the nucleus and identified that its narrow linewidth would present challenges for optical spectroscopy, noting that ``it will be very difficult to find the nuclear level by the method of direct photo excitation using a laser beam. It would take some years for passing the energy interval ... by a laser beam ... because we have to irradiate the target $\approx 100$~s at each step..."~\cite{Tkalya1996}.
He also identified that the unique properties of the isomeric state would allow several new studies, including the ``development of a high stability nuclear source of light for metrology" thereby planting the seed for the development of the nuclear clock.

The following year the first claim of observation of the nuclear isomer decay was reported~\cite{Irwin1997}.
A UV spectrometer was used to measure the optical emission from two powder samples of $\ce{^233U}$, revealing two wide fluorescence peaks around 2.4~eV and 3.5~eV.
Using the fact that roughly 2\% of \ce{^{233}U} decays were expected to populate the $^{229m}$Th state~\cite{Kroger1976}, the latter peak was attributed to isomer photoemission and the former to isomer relaxation by an electron-bridge process involving a low-lying Th electronic state.
This claim was quickly supported by a follow-on experiment, which recorded the UV photoemission spectrum of \ce{^{233}U} dissolved in nitric acid, as well as performed a control experiment using a $^{232}$U sample~\cite{Richardson1998}.
Though the \ce{^{233}U} and \ce{^{232}U} samples produced very similar spectra to each other, which were attributed to liquid scintillation, their subtraction showed remarkable agreement with the result of Irwin and Kim.

Soon after appearing, however, these results were called into question.
An experiment at Lawrence Livermore National Lab prepared an electroplated sample of, presumably, $\ce{^233UO3}$ and observed that photoemission of the sample decreased significantly under vacuum~\cite{Utter1999}.
They noted that the \rev{spectral} features observed by both Irwin and Kim and Richardson et al. were similar to known emission lines of air under a discharge, and concluded the previous measurements had observed $\alpha$-particle induced fluorescence of air.
Almost contemporaneously, an experiment recording the spectrum of photoemission from high-purity \ce{^233UO3} powder was performed at Oak Ridge National Lab~\cite{Shaw1999}.
Using almost three days of integration they resolved several lines around 3.5~eV.
They then recorded the spectrum of an \ce{N2} discharge, which reproduced all but one of the lines observed from the \ce{^233UO3} powder.
They posited the remaining line could be explained by other mundane sources, but left open the possibility that it could be due to \thor{}.
They saw no evidence for the emission around 2.4~eV seen by previous measurements.
Like Utter et al., they concluded that the previous measurements had likely observed $\alpha$-induced air fluorescence.
Concern was also soon expressed based on theoretical grounds.
A comment to Physical Review Letters on the Irwin and Kim result pointed out that neither the 3.5~eV emission nor the 2.4~eV emission attributed to electron bridge decay are expected~\rev{\cite{Karpeshin1999Comment}}.
Instead, the nucleus should decay primarily by electron bridge producing emission around 1.0~eV and 1.8~eV.
In a reply to this comment,~\citet{Irwin1999} agreed that, given the work of Utter et al. and Shaw et al., the 3.5~eV peak was not due to $^{229m}$Th photoemission, but left open the possibility that the red-shifted fluorescence may be due to electron bridge decay.
However, later that year, an experiment performed at Oak Ridge National Laboratory showed that the red-shifted fluorescence could be explained by fluorescence of uranyl ions in the samples~\cite{Young1999}.

Soon after, Tkalya and co-workers analyzed these experiments from a theoretical viewpoint and made several important advances~\cite{Tkalya1999,Tkalya2000}.
First, they pointed out that the `unassigned line' near 3.5~eV in the Shaw et al. work coincided with a known emission line of \ce{N2+}.
Second, they studied the decay of \thorm{} in a metal, showing that it would decay primarily by internal conversion with valence band electrons and exhibit a lifetime of order 1~$\mu$s.
Third they concluded that in a dielectric with a bandgap larger than the isomeric energy, the primary decay mechanism of the isomer would be radiative decay.
They proposed using \ce{ThO2} samples with a synchrotron to excite 29~keV and 72~keV \thor{} lines in order to populate \thorm{}, as well as populating the state directly by a laser.
They envisioned monitoring the excitation by measuring the change in the $\alpha$-decay spectrum predicted for the isomeric state~\cite{Dykhne1996}.

In the following years, progress slowed as attempts were made to understand the photoemission data and identify new routes to measuring the isomer energy.
Several theoretical works~\cite{Band2001, Kalman2001} criticized the comment by Karpeshin and co-workers, arguing that the 3.5~eV emission could be expected and that the electron bridge mechanism should produce light around 2.4~eV.
They also argued that the air fluorescence explanation was not definitive.
However, two measurements were performed that seemed to rule out the possibility of the isomer in the expected energy range.
By quickly separating \thor{} from \ce{^233U} and looking for growth in the $\gamma$-decay spectrum of \thorg{} due to \thorm{}$\rightarrow$\thorg{} decay, scientists at Oak Ridge National Lab constrained that the isomer half-life was either $< 6$~hours or $>20$~days.
Similarly, as described in an unpublished internal report~\cite{Moore2004}, work at Argonne National Lab quickly separated \thor{} from high-purity \ce{^233U} and dissolved it into HCl.
The dissolved solution was placed in an apparatus to record photoemission within 22 minutes of extraction, and the resulting photoemission rate monitored.
No evidence of the isomeric transition was found and it was concluded the isomer half-life was either shorter than 5~minutes or longer than 115~days.
Finally, \rev{\citet{Tkalya2003Optical}} considered the nuclear recoil in uranium compounds and showed the nuclear transition could be completely quenched.
These results, combined with the previous work~\cite{Utter1999,Shaw1999, Young1999}, placed the field at an impasse: $\gamma-$ray spectroscopy suggested a low-energy isomer, but no evidence for the state could be found.

\rev{During these quiet years, \citet{Mitsugashira2003} reported
an $\alpha$-decay signal consistent with \thorm{} and a half-life
of about 14 hours. A follow-up $\alpha$-decay search was reported
by \citet{Kikunaga2005}. In a separate photon-detection experiment,
\citet{Kasamatsu2005} found no signal from thorium in nitric acid.}
Meanwhile, several theoretical studies continued to explore possible
means of nuclear excitation~\cite{Karpeshin2006, Karpeshin2006a}.
One of the most important advances during this time, was the proposal by~\citet{PieTam03} on the use of the isomer as a nuclear clock and to test the variability of the fundamental constants.
In this work, they considered an adaptation of the Dehmelt shelving scheme to allow operation of ion clock based on \ce{^229Th^3+} ions.
While projected clock performance was not calculated they considered mitigation of systematics and suggested a high performing clock could be constructed.
They also suggested that \ce{ThO2}, \ce{ThF4}, and \ce{Th}-doped glasses could be used for nuclear laser spectroscopy and note the system could be used as a short-term frequency reference.
While performance was not investigated, they estimated the temperature sensitivity of the transition at $\delta f/f \sim 10^{-10}$~K$^{-1}$.
Around the same time, a reanalysis of the Hemler and Reich data using updated models of nuclear decay branching ratios and more robust statistical methods found the isomeric energy to be $5.5\pm1.0$~eV~\cite{Guimaraes2005}.
Several experiments also began to continue the search for the isomeric state~\cite{Tordoff2006,Inamura2006}.

{\em The renewed search years: 2007 - 2023.}
The next major development in the field came in 2007 when Livermore and Los Alamos National Labs collaborated with NASA Goddard to use a new high-resolution $\gamma$-ray spectrometer to record $\gamma$-ray spectrometry of $\ce{^233U}$ decay, determining the energy of the isomer as $7.6 \pm 0.5$~eV~\cite{Beck2007}.
From a nuclear physics perspective the shift by a few electron volts was not surprising, because even the new NASA $\gamma$-ray spectrometer had a resolution of $\sim 30$~eV and the isomer energy was inferred from partially overlapping $\gamma$-ray lines.
However, a 7.6~eV photon is squarely in the vacuum ultraviolet (VUV) region of the spectrum, meaning that it is absorbed by even air.
Thus, none of the previous searches for photoemission could have succeeded since the photon would have been absorbed before it reached the detector!
This result re-energized the community, and was the impetus for one of us to begin experiments, to work towards laser excitation of the isomeric transition.

Two years later, a re-analysis of the \citet{Beck2007} data was made by the same team~\cite{Beck2009}.
Because the $\gamma$-ray spectrometers resolution was below than the isomeric energy separation, some transitions were not individually resolved and their center frequency were found by a multi-component fit.
To ascertain the isomeric energy, this fitting required knowledge of the relative magnitude of the overlapping transitions.
By including a cross-band E2 decay from the 42.43-keV $7/2+[633]$ state to the \thorm{} state that had previously been ignored, it was found that the best fit value for the transition energy was instead $7.8\pm0.5$~eV.

During this time, an experiment attempting excitation of the \thor{} in a hollow cathode discharge found a signal that was consistent with an isomer energy between 3~eV and 7~eV and a isomeric state lifetime between 200~s to 600~s \cite{Inamura2009}, though the authors cautioned their results were not conclusive.

The Georgia Tech group then reported the first trapping of \ce{^229Th^3+} ions and demonstrated laser cooling, high-resolution spectroscopy and the extraction of the ground state nuclear moments~\cite{Campbell2009, Campbell2011, SafSafRad2013-Th3plus}, as well as utilization of an E2 electronic transition promising for electron bridge excitation of the nucleus~\cite{Radnaev2012}.
{Working with  Reno and Sydney theory groups, this effort developed the first concrete trapped-ion architecture for a \thor{} nuclear clock, analyzing the electronic structure, excitation pathways, and systematic shifts that would set its performance~\cite{CamRadKuz12}. A central insight was that the high $10^{-19}$ fractional accuracy of such a clock can be attained by choosing stretched hyperfine clock states for which the nuclear and electronic degrees of freedom are largely decoupled.}

Around the same time, work began at PTB to trap \ce{^229Th+} ions towards realizing a clock based on electronic bridge excitation of the nucleus~\cite{Porsev2010}, which has the potential advantage of avoiding the need for a VUV laser system.
Under this effort, \ce{^232Th+} was trapped and spectroscopy was recorded to find potential pathways for electronic bridge excitation~\cite{Herrera2012, Herrera2013}.

In 2008, a new approach to determining the transition and realizing a solid-state clock was begun at UCLA~\cite{Hudson2008, Rellergert2010}.
While previous work~\cite{Tkalya1996, PieTam03} had considered the possibility of driving the optical transition in \ce{ThO2}, \ce{ThF4}, and \ce{Th}-doped glasses, the determination of the isomer energy at $7.8\pm0.5$~eV meant these materials could not be used as they were not transmissive in the VUV~\footnote{As will be discussed later, a decade later \ce{ThF4} was reanalyzed and it is in fact transmissive in the VUV~\cite{Gouder2019}, allowing it be used as host material~\cite{Zhang2024-ThF4}.}.
Therefore, it was proposed that the \thor{} could be doped into high bandgap fluorides, including \ce{LiSrAlF6} and \ce{CaF2}, which can be grown with ultrahigh quality and exhibit electronic bandgaps $>10$~eV, allowing transmission well into the VUV.
This work also performed the first analysis of the potential performance of a solid-state clock showing performance at the $10^{-17}$ level of stability was possible, primarily limited by the temperature dependence of the chemical and electric quadrupole shifts.

This group grew the first \thor{}-doped crystals and performed a synchrotron based search for the isomeric transition~\cite{Rellergert_2010_progress, Jeet2015} in \thor{}:\ce{LiSrAlF6} crystals.
Following the early unsuccessful searchers, the group worked with Tkalya to refine the search region for the transition -- i.e. its possible energy and lifetime -- based on all available nuclear data~\cite{Tkalya2015} and constructed the first VUV laser system for excitation of the nuclear transition~\cite{Jeet2018}.
These experiments were followed by efforts at PTB and Vienna~\cite{Peik2009, Kazakov2012-performance-229Th-clock}, with the TU Wien group developing \thor{}:\ce{CaF2}~\cite{Dessovic2014,Stellmer2015,Stellmer2018}.

During this time, there were a number of attempts made to confirm the existence of the isomeric state and better measure its energy.
At Los Alamos National Lab, recoiling \thor{} produced from $\alpha$-decay of a sample of \ce{^233U} were directed into a \ce{MgF2} plate.
After implantation, these plates were monitored for VUV emission originating from the fraction ($\sim 2\%$) of \thor{} produced in the isomeric state following \ce{^233U} decay.
While these experiments claimed detection of the isomer decay and a measurement of its radiative half life at $6\pm1$~hr, it was immediately met with skepticism~\cite{Peik2013} as radioluminescence from daughter nuclei easily explained the signal.
Further, it was expected that non-radiative decay of the isomer likely occurred during and after the violent implantation process~\cite{Barker2018}.
As will be discussed later, it is now known the isomeric state lifetime in high band gap fluorides is of order of several 100~s, thus it can be concluded these experiments did not observe isomeric decay.
Several attempts were also made at Lawrence Livermore National Lab to measure the isomeric state via both internal conversion and radiative decay~\cite{Swanberg2012,Ponce2018}.
No clear signal was observed and attributed to an internal conversion lifetime that was shorter than the measurement timescales.
At UCLA, in collaboration with the National Institute of Standards and Technology (NIST), \thor{} produced from $\alpha$-decay of a sample of \ce{^233U} was directed onto a superconducting nanowire detector~\cite{Jeet2018}.
The experiment searched for the transition by recording instances of two detection events within $30~\mu$s -- the detector registers the implantation of \thorm{} and then again when it decays via internal conversion.
Using the imaging capability to measure only double events that occurred at the same place within the nanowire, a signal consistent with the \thorm{} was observed, however, it could not be confirmed as the signal could also be explained by the decay of a \thor{} daughter, \ce{^213Po}~\cite{ElwellPhD2024}.

Experiments were also begun at Ludwig Maximilian University to develop a buffer gas stopping cell for deceleration and guiding, through a quadrupole mass filter, of recoiling \thor{} produced from $\alpha$-decay of a thin sample of \ce{^233U}~\cite{vonderWense2015}.
These experiments aimed to use the fraction of \thor{} ($\sim 2\%$) produced in the isomeric state following \ce{^233U} decay as a means to measure the energy of the \thorm{} state, either by detection of radiative relaxation of the isomer~\cite{Seiferle2016photo} or measuring the energy released from e.g. internal conversion.
In 2016, this group realized the first direct detection of the isomeric state decay using this apparatus~\cite{VonderWense2016a} by detecting the internal conversion decay of \thorm{} after ``soft landing'' \ce{Th^n+} ions onto a multi-channel plate detector.
They were also able to constrain the isomeric state energy to be between 6.3~eV and 18~eV, the isomeric lifetime to be $>60$~s, and measured the internal conversion lifetime to be $10\pm1~\mu$s~\cite{Seiferle2017lifetime}, which was consistent with an earlier prediction~\cite{Tkalya2015}.
They later modified the experimental apparatus to measure the energy of the internal conversion electrons~\cite{Seiferle2017} and inferred that, subject to corrections due to atomic structure, the isomer energy was $8.28\pm0.17$~eV~\cite{Seiferle2019}.
Using the same apparatus, laser spectroscopy was performed that recorded the hyperfine structure of both the ground and isomeric states~\cite{Thielking2018}, providing a direct measurement of the isomeric state nuclear moments\cite{Thielking2018, Muller2018}.
Finally, using this apparatus with the superconducting nanowires developed for the UCLA-NIST experiment, a signal consistent with an isomer energy of around $8.6\pm 0.2_{\textrm{stat}} \pm 0.1_{\textrm{sys}}$~eV was measured~\cite{Oneil2025}.

Meanwhile, a collaboration at the SPring-8 facility demonstrated resonant excitation of the cross-band transition at 29~keV between the nuclear ground state of \thor{} and the second excited state, corresponding to the first excited state of the $(3/2)^+[631]$ rotation band~\cite{Masuda2019XrayPumpingTh229}, in a \rev{thorium target deposited on graphite}.
This experiment represented the first demonstrated optical pumping of \thor{} into the isomeric state, and enabled the first independent measurement of the branching ratio from the 29 keV second excited state to the ground and isomeric state since the Beck et. al. measurements~\citet{Beck2007}.
The measured value of the branching ratio was $1/(9.4\pm 2.4)$, consistent with the Beck et al. value and within the range of values given in~\citet{Tkalya2015}.
\rev{The measurement found the isomer energy to be in the range of 2.5 - 8.9~eV.}

The next major breakthrough came in 2022, when an experiment performed at the CERN ISOLDE facility observed radiative decay of \thorm{}~\cite{Kraemer2022a}.
In this experiment, \ce{^229Fr} and \ce{^229Ra} were implanted into \ce{MgF2} and \ce{CaF2} samples at ISOLDE, which in the $\beta$-decay chain feed into $\ce{^229Ac}$.
When \ce{^229Ac} itself $\beta$-decays, with a half life of roughly 1 h, it produces a significant fraction ($>10\%$) of \thor{} in the \thorm{} state.
The radiative decay of the \thorm{} was observed using a VUV spectrometer and the isomeric state energy determined to be $8.338\pm 0.024$~eV and its half-life when embedded in \ce{MgF2} was measured to be $670 \pm 102$~s~\cite{Kraemer2022a}.
This experiment was similar to the earlier unsuccessful experiments at Los Alamos National Lab~\cite{Zhao2012}, which attempted to implant \thorm{} into \ce{MgF2}.
The chief advantage of the ISOLDE experiment appears to be that \thorm{} is produced in the comparatively gentle $\beta$-decay process well \emph{after} the parent ion is already implanted in the crystal.
This presumably allows the \thorm{} to be in an electronic environment, as discussed later, that does not provide substantial non-radiative relaxation pathways.

\begin{figure}[th]
    \centering
    \includegraphics[width=0.95\columnwidth]{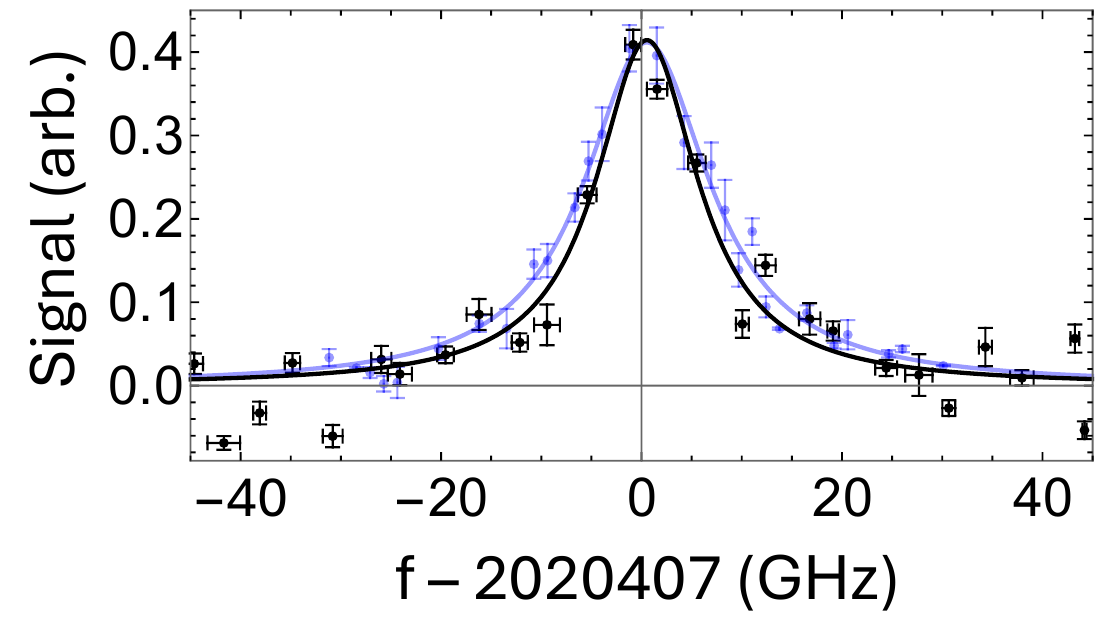}
    \caption{(Color online) Nuclear spectroscopy of the
    \thor{} transition in \thlisaf{}  and \ce{ThO2}.
    Lighter points and curve (blue online) show the
    radiative-decay spectrum measured in \thlisaf{}.
    The black points show the laser-based
    conversion-electron M\"ossbauer spectrum of a thin
    $^{229}\mathrm{ThO}_2$ target.  Adapted from ~\cite{Elwell2025-ThO2}}
    \label{fig:cems-spectrum}
\end{figure}

{\em \rev{Direct laser excitation and beyond: 2024 onward.}}
While the importance of the ISOLDE experiment cannot be overstated as it provided strong proof of the existence of the isomeric state and the first determination of its energy that did not require significant correction and calibration effort, it still left a relatively large range -- about a 2~nm wide band centered at 148.7~nm -- to be explored before laser excitation could be achieved.
Therefore, immediately after the result was announced groups around the world began efforts to excite the \thor{} nucleus with a laser.
The PTB/Vienna group reported the first laser excitation~\cite{Tiedau2024-caf2} using \thor{}:\ce{CaF2}, with the UCLA group reporting excitation in \thor{}:\ce{LiSrAlF6} shortly thereafter~\cite{Elwell2024-lisaf}.
The measured nuclear excitation spectrum in \thlisaf{} is shown in Fig.~\ref{fig:cems-spectrum}.
Both groups observed a narrowband nuclear spectrum, limited by their laser linewidths, which was not present in \regthor-doped versions of their crystals.
The PTB/Vienna group measured the central nuclear transition and lifetime in the crystal as $2020409\pm 7$~GHz and $630\pm15$~s, respectively, while the UCLA team measured the transition at $2020407.3 \pm 0.5_{\textrm{stat}} \pm 3_{\textrm{sys}}$~GHz and lifetime in the crystal as $568\pm 13_{\textrm{stat}} \pm 20_{\textrm{sys}}$~s.
Given that both groups using two different hosts measured an identical transition frequency within error and neither group saw a signal from \regthor{}-doped crystals, it was clear that the dream first identified by Kroger and Reich in 1976 had finally been realized 48 years later.

Almost simultaneously with the laser excitations in crystals, a group in Tokyo reported results of trapping \thorm{} in an ion trap~\cite{Yamaguchi2024}.
Like earlier work trapping \thorm{} produced in \ce{^233U} decay~\cite{Thielking2018}, they were able to measure hyperfine-resolved electronic spectroscopy of \thorm{}, but thanks to their long trapping lifetimes they were also able to use this spectroscopy to monitor the radiative decay of the isomeric state.
This was the first measurement of isomeric decay in an ion, finding a lifetime of
$2000^{+900}_{-400}\, \mathrm{s}$.
Since the M1 lifetime in the crystal is shortened by a factor of $n^3$, where $n$ is the index of refraction for a photon resonant with the isomeric transition~\cite{Tkalya2000-M1inMedium,Nienhuis1976a}, the measurement in \thor{}:\ce{CaF2} suggests an isomer decay lifetime of $2510\pm70$~s~\cite{Tiedau2024-caf2}.
The measurement in \thor{}:\ce{LiSrAlF6} is complicated by the fact the index of refraction is, as of yet, unmeasured; however, a preliminary estimate suggests a lifetime of $1878\pm50$~s.
These values are all in rough agreement with an earlier estimate of $2450\pm1320$~s using the Alaga rules in conjunction with existing $\gamma$-ray spectroscopy data~\cite{Tkalya2015}.

The Okayama group also followed up on their earlier demonstration of x-ray excitation, by populating the isomeric state by the 29~keV line and observing fluorescence from the isomeric state~\cite{Hiraki2024}.
They measured the isomeric transition at \rev{$8.367 \pm 0.024$}~eV
and an isomeric state lifetime of $645 \pm 36$~s, and observed that
the x-ray beam led to quenching of the isomeric state.

A few months later, using \thor{}:\ce{CaF2} crystals produced in Vienna, a team at the University of Colorado used a narrow-band VUV frequency comb to resolve the individual hyperfine components of the transition, measuring the transition frequency as $2,020,407,384, 335 \pm 2$~kHz.
With hyperfine-resolved spectroscopy they were able to extract the electric field gradient due to the crystal experienced by the \thor{}, and showed that the original (controversial) prediction that \thor{} would predominantly substitute into one configuration in the crystal was correct.
This same experiment was also able to measure the nuclear transition frequency as a function of temperature, finding a dependence of the transition frequency due to both the electric quadrupole interaction and the chemical shift~\cite{Higgins2025-Temperature-dependence-ThCaF2} on the order of 0.1-1~kHz/K, in line with earlier predictions~\cite{Rellergert2010}.

Next the same group examined the reproducibility of the transition frequency, finding that one transition is first order insensitive to the temperature at $196\pm5$~K and was unchanged at the $10^{-13}$ level over 7 months~\cite{Ooi2026}.
They also observed that the transition linewidth is dependent on the
\thor{} concentration in the crystal, \rev{with an extrapolated linewidth
of $\sim25$~kHz at zero dopant density, possibly arising from
host-crystal defects unrelated to \thor{}}.
This concentration dependence connects the discovery history directly to the materials discussion below, where magnetic and defect-mediated broadening become central clock-design questions.

An experiment using crystals from the same batch, grown by the Vienna group, was also performed by the Okayama group using a laser with moderate linewidth, but higher spectral density allowing more sensitive observation of the \thor{}~\cite{Hiraki2026Science}.
This experiment revealed several populations of \thor{} occupying different configurational sites in the \ce{CaF2} crystal.
Their electric field gradients were used to distinguish the populations and doping configurations were postulated to explain them; interestingly, the most common site identified by the Okayama group was absent in the earlier spectroscopy from the Colorado group.

The next important development came from work towards understanding the coupling of the nuclear and electronic degrees of freedom in solid-state hosts.
The experimental effort towards nuclear excitation in \thor{}:\ce{LiSrAlF6} had reported for several years a few seconds timescale fluorescence signal near the isomeric transition energy~\cite{Elwell2024-lisaf}.
Once the nuclear transition was found, it also became clear that in some crystals not all the \thor{} participated in the long lifetime signal.
These facts were explained by an electronic-defect-based quenching enabled by the hyperfine coupling of the electronic and nuclear degrees of freedom~\cite{Elwell2024-lisaf} and later developed into a full model of internal conversion in a solid-state environment~\cite{Morgan2025_internal_conversion}.
This model was extended to include the effects of photo-excited electronic defects and was experimentally verified~\cite{Terhune2025-photo-induced} in \thor{}:\ce{LiSrAlF6} with similar results shown in \thor{}:\ce{CaF2}~\cite{Schaden2024-Th-quenching}.
\rev{Recent theoretical work has analyzed the host chemical shift,
predicting an approximately 80~MHz-wide frequency window for the
solid-state hosts considered, whereas the corresponding transition
frequencies in \Thppp{} and \Thpppp{} differ from the central frequency
in \thcaf{} by hundreds of megahertz}~\cite{Perera2025PRL}.

 An important advance came when it was demonstrated that the isomeric state could be excited in a material {\em opaque} to the 148.4~nm light required for direct nuclear excitation~\cite{Elwell2025-ThO2}. This result opened the possibility of using \thor{} in a much broader class of materials. In this work, a thin layer of \ce{ThO2} was electro-deposited onto a stainless steel disk.
Though \ce{ThO2} strongly absorbs the VUV laser, \thor{} in the first $\sim 20$~nm of the target can be excited.
These excited nuclei then decay by the internal conversion channel and can eject an electron from the material, which can be recorded.
Using this system, the transition was measured to be at $2,020,407.5 \pm 2_{\textrm{stat}} \pm 30_{\textrm{sys}}$~GHz  and the internal conversion lifetime was found to be $\sim 10~\mu$s, opening the door to a nuclear clock based on an electrical current readout.
The resulting first ever laser conversion-electron M\"ossbauer spectrum is shown in Fig.~\ref{fig:cems-spectrum}.
A theoretical formalism was developed for calculating internal conversion decay channels in materials and found reasonable agreement with experimental observations.

\rev{Finally, narrow-linewidth continuous-wave (CW) VUV sources have enabled absorption spectroscopy and clock operation in \thcaf{}.
\citet{Xiao2026} demonstrated a CW source at 148.4~nm based on four-wave mixing in cadmium vapor.
\citet{Morawetz2026Nature} demonstrated CW nuclear absorption spectroscopy, which avoids waiting for the isomer fluorescence decay and permits rapid generation of a clock error signal.
\citet{Huang2026_nuclear_clock} used frequency-modulation absorption readout to lock a CW VUV laser to a resolved nuclear transition in \thcaf{}, reporting an in-loop fractional frequency instability of $5\times10^{-13}$ at 1~s averaging time.
An independent clock implementation reported a fractional frequency instability of $3\times10^{-12}$ at 1~s~\cite{decol2026thorium229opticalclock}.}

\section{Clock basics}
\label{Sec:ClockBasics}

Telling time relies on counting cycles of a periodic process.
A clock is useful only insofar as it realizes an oscillation period or frequency that is stable, reproducible, and known.
In a quantum system, this frequency is set by the Bohr relation, $\nu_0 = (E_e-E_g)/{h}$, for a transition between stationary states $\ket{g}$ and $\ket{e}$.
In an optical clock, a tunable laser is locked to this transition, and time is inferred by counting the cycles of the stabilized laser field.

The practical clock problem is therefore to make the realized transition frequency stable and reproducible.
Any environmental perturbation that shifts the transition frequency also changes the clock period.
For this reason, frequency shifts due to host materials, defects, and charge states are not secondary engineering details, but central elements of clock design.
Reproducibility is equally important: independent realizations of the same clock design should yield the same transition frequency within their stated uncertainties.
In this limit, the clock can serve as a frequency reference or standard.

The performance of a clock is usually separated into two questions.
Its \textit{accuracy} is determined by how well perturbative shifts of the realized frequency can be calculated, measured, or canceled.
Its \textit{stability} is determined by the rate at which repeated interrogations average down the statistical uncertainty.
In the projection-noise-limited regime characteristic of quantum measurements, the fractional frequency instability may be written schematically as
\begin{equation}
    \sigma_y(\tau) \sim
    \frac{\chi}{Q_{\rm eff}\sqrt{N}}
    \sqrt{\frac{T_c}{\tau}},
    \label{eq:clock-stability-scaling}
\end{equation}
where $Q_{\rm eff}=\nu_0/\Delta\nu_{\rm eff}$ is the quality factor of the interrogated resonance, $N$ is the number of interrogated nuclei, $T_c$ is the experimental cycle time, $\tau$ is the averaging time, and $\chi$ is an order-unity factor that depends on the interrogation protocol, line shape, state preparation, and readout efficiency.
As discussed in Sec.~\ref{Sec:History}, the exceptional feature of \thor{} is that its first nuclear excited state lies in the VUV, while retaining a sub-mHz natural linewidth and hence a natural quality factor of order $Q_{\rm nat}\sim 10^{19}$.
In practice $Q_{\rm eff}$ is set by the realized interrogation time, laser coherence, and environmental broadening, but the enormous $Q_{\rm nat}$ provides the headroom for an exceptionally narrow clock resonance.
Moreover, because the transition can be driven in a solid-state host, a \thor{} nuclear clock can in principle interrogate a macroscopic ensemble of nuclei, increasing $N$ far beyond what is available in a trapped-ion clock and correspondingly reducing the projection-noise-limited instability.

Optical atomic clocks have reached extraordinary performance by exploiting narrow electronic transitions while controlling environmental shifts at the $<10^{-18}$ level and below~\cite{LudBoyYe15-OpticalClocks-review}.
A nuclear clock follows the same metrological logic, but replaces the electronic reference with a transition internal to the nucleus~\cite{PieTam03}.
At first sight, this substitution seems to promise exceptional environmental immunity: the small nuclear size suppresses electric multipole couplings, while nuclear magnetic moments are reduced by the nucleon mass.
{This intuitive expectation is misleading, as the hyperfine interaction couples nucleonic and electronic degrees of freedom~\cite{Rellergert2010,CamRadKuz12}}.
Thus, unless special care is taken, a nuclear clock is no different than an atomic clock, as the hyperfine coupling ensures the nuclear clock inherits the same electronic sensitivity to perturbation.
Perturbations that act primarily on the electronic, motional, or crystal degrees of freedom can thus be transferred to the nuclear transition, see Sec.~\ref{Sec:El-Nucl-coupled}, so the intrinsic robustness of the bare nucleus does not by itself guarantee clock accuracy.
A central task is to then identify clock states, host environments, and techniques in which these couplings are common-mode, averaged away, or intrinsically small~\cite{CamRadKuz12,Beloy2023a,Morgan2025-design-of-Th-materials,morgan2025-spinless}.

Thus, in a properly designed platform, the \thor{} nuclear isomeric transition combines reduced environmental sensitivity and increased fundamental-physics sensitivity with an excitation energy accessible to modern laser spectroscopy.
This accidental gift of nuclear structure makes possible the two principal clock architectures~\cite{PieTam03} discussed in this review.
Trapped-ion nuclear clocks, Sec.~\ref{Sec:Ions}, emphasize controlled systematics and high accuracy~\cite{CamRadKuz12}.  Solid-state nuclear clocks, Sec.~\ref{Sec:SolidState}, interrogate many \thor{} nuclei in a crystal, gaining an ensemble signal that can average down rapidly, offering a superior stability~\cite{Rellergert2010}.

\section{The coupled electron--nucleus system}
\label{Sec:El-Nucl-coupled}

As emphasized in the previous section, interactions between the nucleus and its surroundings are central to both nuclear-clock accuracy and stability.
They may either perturb the clock transition or provide the mechanism by which it is addressed, controlled, and detected.
We organize the relevant physics in terms of a schematic compound-system Hamiltonian,
\begin{equation}
    H^{(p)} =
    H_{\rm nuc}+H_{\rm el}^{(p)}+H_{\rm lat}^{(p)}
    +H_{\rm int}^{(p)}+H_{\rm ext}^{(p)} + \ldots\,.
    \label{eq:compound-clock-hamiltonian}
\end{equation}
Here $p$ denotes the ion or material platform.
The term $H_{\rm nuc}$ is the bare nuclear Hamiltonian.
The electronic Hamiltonian $H_{\rm el}^{(p)}$ describes the electrons in the reference Coulomb field of the \thor{} nucleus and in any platform-dependent electronic environment, such as trapping fields or crystal fields.
The term $H_{\rm lat}^{(p)}$ denotes host degrees of freedom, including phonons, defects, strain, and, for trapped ions, the external motional environment.
Couplings among the nuclear, electronic, and host degrees of freedom are collected in $H_{\rm int}^{(p)}$; these include the state-dependent Coulomb interaction with the finite nuclear charge distributions of the two nuclear states, as well as hyperfine interactions such as magnetic-dipole and electric-quadrupole couplings.
Finally, $H_{\rm ext}^{(p)}$ contains applied and ambient fields, including probe lasers, magnetic fields, trapping fields, blackbody radiation, vacuum fluctuations of electromagnetic fields, and technical perturbations.
The separation in Eq.~\eqref{eq:compound-clock-hamiltonian} is schematic rather than unique: applied fields may be screened or modified by the environment, and perturbations that act first on the electronic or lattice degrees of freedom can be transferred to the nucleus through $H_{\rm int}^{(p)}$.
Reducing this complex many-body problem to its essential terms is platform and application specific, and remains an active part of nuclear-clock modeling.

\begin{figure}[ht]
    \centering
    \includegraphics[width=0.99\columnwidth]{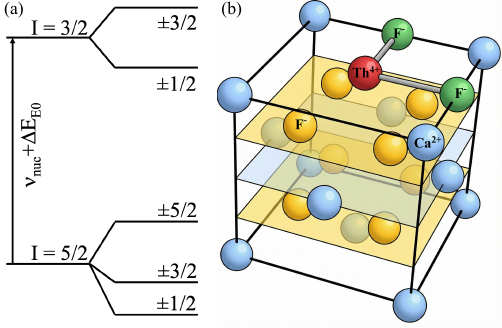}
\caption{(a) The local crystal electric-field gradient splits the
$I=5/2$ ground-state and $I=3/2$ isomeric manifolds into Kramers
doublets, labeled by their dominant $\ket{I,m_I}$ components.
\rev{(b) Idealized schematic of one possible local structure of thorium-doped \caf{}.
Host \ce{Ca^2+} ions are blue and host \ce{F^-} ions are yellow.
The red Th$^{4+}$ dopant replaces a \ce{Ca^2+} ion at the center of the upper cell face.
Two additional interstitial fluoride ions (green) compensate the excess charge.
This microscopic defect complex sets the local electric-field gradient responsible for the nuclear splitting in panel (a), as well as the isomer shift and electronic defect spectrum.
Adapted from \citet{Dessovic2014}.
Panel (b) was prepared with Google Gemini; the panels were assembled with Codex-assisted code.}}
    \label{fig:HE2_defect}
\end{figure}

Experiments do not measure transition frequencies of the nuclear Hamiltonian alone.
Rather, they measure energy differences between eigenstates of the compound system described by the Hamiltonian~\eqref{eq:compound-clock-hamiltonian}.
The phrase ``nuclear transition frequency'' remains useful shorthand, but the realized clock frequency is platform dependent because the nuclear states are coupled to the electronic cloud and, in a solid, also to the lattice.
Moreover, a single bare nuclear transition can appear as several resolved frequencies.
For example, $J\neq0$ electronic levels in an ion split into nuclear-state-dependent hyperfine manifolds, while in a crystal the electric-field gradients split the nuclear sublevels into Kramers doublets, as shown in Fig.~\ref{fig:HE2_defect}(a).
Even after the hyperfine or crystal-field structure is removed, the remaining ``central'' transition frequency is still platform dependent.
For example, the transition energy measured in \thcaf{} is $8.3557\ldots\,\mathrm{eV}$~\cite{Zhang2024-Th229Comb}, whereas the corresponding bare-nucleus energy is inferred to be $8.272(22)\,\mathrm{eV}$~\cite{Perera2025PRL}; see Table~\ref{Tab:Th229m-props}.
This difference arises because the electronic energy depends on the distinct nuclear charge distributions of the ground and isomeric states, leading to the so-called isomer, or chemical, shift.

The coupling term $H_{\rm int}^{(p)}$ can be separated into two classes of diagrams that are either  ``diagonal'' or ``off-diagonal'' in the nuclear-state basis.
The diagonal diagrams do not change the state of the nucleus, while the off-diagonal do.
We mainly focus on  the electron--nucleus interaction.

The {\em diagonal} part of the electron--nucleus interaction produces several  distinct level shifts and manifolds.
We distinguish between various multipoles of a virtual photon: E0, M1, E2, M3, and E4 relevant to \thor{}.
The nuclear vertex is characterized by various parity-even form-factors or nuclear moments, with known moments listed in Table~\ref{Tab:Th229m-props}.
The E0 form factor is characterized by the r.m.s radii of the nuclear charge distributions - it leads to the isomer shift.
The higher multipoles produce hyperfine structure when the electronic or crystal environment supplies the fields at the nucleus.

The hyperfine manifolds generated by the two nuclear states differ.
For \ce{Th^3+} in its ground electronic $5f_{5/2}$ state, $J=5/2$.
When the \thor{} nucleus is in the ground state, $I_g=5/2$, the allowed total angular momenta are
$F=0,1,\ldots,5$, giving six hyperfine levels.
When the nucleus is in the isomeric state, $I_m=3/2$, the allowed values are
$F=1,2,3,4$, giving four hyperfine levels.
\rev{An early discussion of identifying the nuclear isomer through changes in the electronic spectrum was given by \citet{Karpeshin1996}.
In a double-resonance detection scheme, one drives the nuclear transition and then probes an electronic transition whose hyperfine pattern depends on whether the nucleus is in its ground or isomeric state~\cite{PieTam03,CamRadKuz12}.}
Similar ideas apply to solids, as the nuclear electric quadrupole (E2) moments couple to the local electric-field gradients, splitting the hyperfine manifold in a crystal, see Fig.~\ref{fig:HE2_defect}.
These ``diagonal'' effects are valuable diagnostics of the nuclear and material state, but they also change the clock frequency from the bare-nucleus value and can broaden the nuclear line, see Sec.~\ref{Sec:SolidState}.

\begin{table*}[th!]
\caption{\label{tab:electron-mediated-processes}
Electron-mediated excitation and decay processes relevant to the \thor{} isomer.}
\begin{ruledtabular}
\begin{tabular}{p{0.16\textwidth}p{0.11\textwidth}p{0.43\textwidth}p{0.22\textwidth}}
Process & Direction & Mechanism & Representative references \\
\hline
Direct radiative coupling
& Excitation or decay
& Nuclear absorption or emission of a real VUV photon; electrons modify only the optical environment, such as the refractive index or mode density.
& \cite{Nienhuis1976a,Tkalya2001_purcell} \\

Internal conversion (IC)
& Decay
& The excited nucleus ejects a bound electron into the continuum; the channel is open only when $\hbar\omega_{\rm nuc}$ exceeds the electron binding energy.
& \cite{Karpeshin2007,TkalyaSi2020-Th229anion} \\

Bound IC
& Decay
& The excited nucleus promotes a bound electron to an unoccupied bound state, leaving the atom or ion electronically excited rather than ionized.
& \cite{Band2001,Karpeshin2007} \\

Band-to-defect IC
& Decay
& A valence-band electron is promoted into an intragap defect state, leaving a hole in the valence band.
& \cite{Morgan2025_internal_conversion} \\

Band-to-band IC
& Decay
& An occupied band electron is promoted to a higher-energy band in low-gap hosts.
& \cite{Elwell2025-ThO2} \\

Nuclear excitation by electron transition (NEET)
& Excitation
& A resonant bound--bound electronic transition transfers its energy to the nucleus.
& \cite{Karpeshin1996} \\

Nuclear excitation by electron capture (NEEC)
& Excitation
& A continuum electron is captured into a bound orbital while exciting the nucleus.
& \cite{Palffy2006,Zhang2023-NEEC-etc,Xu2026NEECions} \\

Nuclear excitation by inelastic electron scattering (NEIES)
& Excitation
& An inelastic collision with a free, plasma, or conduction electron excites the nucleus.
& \cite{Tkalya1999,Tkalya2000,Zhang2023-NEEC-etc} \\

Electronic bridge (EB)
& Excitation or decay
& Nuclear and electronic transitions are coupled by off-diagonal electron--nucleus interactions, while photons supply or remove the energy mismatch.
& \cite{Tkalya1992,Kalman2001,Porsev2010,Nickerson2021} \\

Defect-mediated or photo-induced quenching
& Decay
& Defect, band, or photo-excited states open IC- or EB-like electronic decay pathways.
& \cite{Terhune2025-photo-induced,Schaden2024-Th-quenching,guan2026_xray_quenching_caf2}
\end{tabular}
\end{ruledtabular}
\end{table*}

The {\em off-diagonal} part of the same $H_{\rm int}^{(p)}$ coupling determines how energy moves between the nucleus and the electronic or material environment.
The main cases used in the \thor{} literature are detailed in Table~\ref{tab:electron-mediated-processes}.
They differ mainly in the direction of energy flow and in the electronic boundary conditions.
In internal conversion (IC), an initially excited nucleus transfers its energy to the electrons; in the time-reversed or inverse-conversion direction, processes such as nuclear excitation by electron transition (NEET), nuclear excitation by electron capture (NEEC), and inelastic electron scattering can pump ground state nuclei into the isomeric state.

IC in the neutral thorium atom provides the simplest energy-conservation example.
Ordinary bound-to-continuum IC is open because the isomer energy, $\hbar\omega_{\rm nuc}\simeq8.36$~eV, exceeds the first ionization potential of Th, $\mathrm{IP}=6.3067(2)$~eV~\cite{Kohler1997}.
The final state is therefore $\ce{Th^+}+e^-$, with the conversion-electron kinetic energy set by
$
    \varepsilon_f =
    \hbar\omega_{\rm nuc}
    - \mathrm{IP},
$
where any excitation energy left in the residual ion and the small recoil energy have been suppressed.
This channel was predicted to shorten the lifetime of neutral \thorm{} to the microsecond scale\rev{~\cite{Strizhov1991,Karpeshin2007,Tkalya2015}}.
Experiments with extracted \thorm{} ions then observed the IC signal after surface neutralization, and measured a neutralization-triggered IC half-life of $7(1)~\mu$s~\cite{VonderWense2016a,Seiferle2017lifetime}.
By contrast, the IC channel is energetically forbidden in \ce{Th} ions for which the relevant $\mathrm{IP}>\hbar\omega_{\rm nuc}$.

As for the solid-state \thor{} platforms, in an ideal wide-band-gap ($E_\mathrm{gap} > \hbar\omega_{\rm nuc}$) crystal, such as \ce{ThF4}, IC is energetically forbidden.
In a doped high-bandgap insulator, however, a localized electronic level in the gap can reopen the IC channel: the nucleus quickly de-excites while an electron is promoted from the valence band or from a defect state to a higher electronic state~\cite{Morgan2025_internal_conversion,Terhune2025-photo-induced}.
In a solid, both electronic and phononic degrees of freedom may participate in the IC process~\cite{Morgan2025_internal_conversion}.
In low-band-gap crystals, such as \ce{ThO2}, the IC promotes valence band electrons into the conduction band, a process used to read out the nuclear excitation~\cite{Elwell2025-ThO2}, see conversion-electron spectrum in Fig.~\ref{fig:cems-spectrum}.

Finally, an electronic bridge (EB) provides an indirect nuclear excitation or decay pathway in which the nuclear transition is coupled to an electronic transition by the off-diagonal electron--nucleus, or hyperfine, interaction.
In excitation, applied light drives the electronic shell while hyperfine coupling transfers amplitude to the nucleus; in decay, the time-reversed process allows the nuclear energy to be released through the electronic environment rather than by direct VUV photon emission~\cite{Tkalya1990,Tkalya1992,Porsev2010,Nickerson2021}.
\rev{\citet{Wang2024HEB} modeled two-photon EB excitation in
\Thppp{} using quantum master equation
and predicted a
near-complete transfer to the isomer state through stimulated Raman
adiabatic passage, which suppresses population of the lossy
intermediate electronic state.}
Related nonlinear excitation schemes have also been proposed for solid-state \thor{} systems.
For example, \citet{Xu2023a} suggested that the optonuclear quadrupolar effect could enable two-photon pumping in wide-band-gap thorium compounds, potentially populating the isomer and producing population inversion in a solid-state gain medium.

A crystal offers a large ensemble of nuclei, but it also bathes the nucleus in phonons and crystal fields, further modified by impurities and defects introduced either during growth or by radiation-induced damage.
Some of these perturbations produce static shifts; others broaden or quench the transition dynamically.
Recent experiments and theory show that defect-assisted and photo-induced quenching can be understood as nuclear relaxation mediated by electronic states inside the band gap~\cite{Elwell2024-lisaf,Terhune2025-photo-induced, Schaden2024-Th-quenching,guan2026_xray_quenching_caf2}.
If controlled, this can improve the clock duty cycle in some implementations.
The design problem for a solid-state nuclear clock is therefore closer to a combined nuclear, atomic, and condensed-matter problem than to a purely nuclear spectroscopy problem.

To summarize, nuclear structure fixes the existence, multipolarity, and intrinsic radiative strength of the \thor{} isomer, see Table~\ref{Tab:Th229m-props}. Atomic and material structure determine the observable frequency, linewidth, excitation pathway, and additional non-radiative decay channels.
A useful nuclear clock must control all of these ingredients at once.

\section{Ion Nuclear Clocks}
\label{Sec:Ions}

As discussed in Sec.~\ref{Sec:El-Nucl-coupled}, the  isomer state in
neutral \thor{} decays primarily by internal conversion (IC), ejecting an electron.
For ground-state \thor{} ions, ordinary IC  is energetically forbidden.
Then the isomer can  decay radiatively on the natural nuclear timescale, giving access to the extremely narrow intrinsic linewidth.    The trapped-ion implementation is therefore the most direct realization of a \thor{} nuclear clock: the host environment is not a crystal,
but an atomic ion whose electronic structure can be prepared, cooled, and interrogated using the tools of precision ion spectroscopy.
The original proposal~\cite{PieTam03} emphasized this point for laser-cooled $^{229}\mathrm{Th}^{3+}$ in a radiofrequency trap: if the nuclear transition can be addressed in a trapped ion, the clock can combine the systematic control of optical ion clocks with a reference transition that is primarily nuclear rather than electronic.

\citet{CamRadKuz12} sharpened this idea by identifying a particularly
favorable choice of clock states in \ce{^{229}Th^3+}. This ion has a single valence electron orbiting a tightly-bound radon-like \ce{Th^4+} closed-shell  core.
They proposed using the stretched hyperfine states
within the electronic ground level $5f\,{}^2F_{5/2}$ in the nuclear ground and
isomeric manifolds,
$|5f_{5/2},I_g=5/2,F=5,m_F=\pm5\rangle$ and
$|5f_{5/2},I_m=3/2,F=4,m_F=\pm4\rangle$.
 Averaging the two opposite
stretched-state transitions defines a virtual clock transition.  These
stretched states factorize into a product of the same electronic wave function and
the nuclear states. Then a perturbation that acts mainly on the electronic cloud shifts the two clock levels nearly in common mode and cancels in the transition frequency.
 This is not the case of any other hyperfine state.
The remaining differential shifts arise
only from the difference in nuclear moments of the two nuclear states, finite nuclear-size effects, and
hyperfine-mediated electronic response, resulting in a projected $10^{-19}$ accuracy of such a nuclear clock.

\citet{Beloy2023a} showed that the parasitic rf magnetic field associated with ion trapping can induce an ac Zeeman shift that may become a dominant systematic in the \Thppp{} clock.
He proposed suppressing this shift either by reducing the rf magnetic field at the ion or by operating at a dc bias field where the differential ac-Zeeman coefficient of the averaged clock transition vanishes.
\cite{ZahMatHume2025-ThDM} proposed a two-photon EB excitation scheme in \Thppp{} that replaces direct narrowband $148$~nm interrogation with UV fields near $270$ and $330$~nm.

Modern ion-clock proposals now span several thorium charge states rather than only \Thppp{}.
Some schemes use the electron shell as an excitation resource.
In \ce{^{229}Th^{2+}}, \citet{Yudin2025a} proposed two-photon spectroscopy of the nuclear transition with a monochromatic field near $296.76$~nm, using an EB enhancement from a nearby electronic level.
Similarly, \citet{Dzuba2025BridgeThII} identified  EB pathways in \ce{Th+}, where the dense electronic spectrum can strongly enhance nuclear excitation.
These approaches may relax the requirement of direct narrowband VUV radiation at $148$~nm, but the same electron--nucleus coupling that enables the enhancement also makes the excitation rate and isomer lifetime sensitive to detailed atomic structure.

A complementary strategy is to use an electronic state with total angular momentum $J=0$ so the nuclear and electronic wavefunctions are decoupled.
Following \citet{Rellergert2010}, \citet{Flambaum2025} proposed exploiting this property in \Thpppp{}, whose closed-shell ground state has $J=0$ and therefore suppresses blackbody-radiation and stray-field shifts mediated by the electrons.
The trade-off is experimental: \Thpppp{} lacks convenient optical resonance lines for direct laser cooling and state detection, so sympathetic cooling and quantum-logic readout would likely be required.
\citet{Maurya2026PRA} recently proposed an analogous hyperfine-free route in \ce{^{229}Th^{2+}} based on the metastable $6d^2\,{}^3P_0$ level.
This state has $J=0$ and a radiative lifetime of $\sim 100$~s.
Compared with \Thpppp{}, the \ce{Th^{2+}} proposal offers more accessible optical manipulation.
The nuclear Zeeman structure remains for both \ce{^{229}Th^{2+}} and \ce{^{229}Th^{4+}} ions, so residual first-order magnetic-field shifts would still have to be removed by averaging over opposite Zeeman components.

While recent ion-trap experiments on \thorm{} have already used hyperfine-resolved electronic spectroscopy to measure the isomer properties~\cite{Thielking2018} and monitor the radiative decay of the isomer~\rev{\cite{Yamaguchi2024}},
at present, the nuclear clock transition frequency has only been measured in solid-state hosts, and not yet in trapped ions.
\citet{Perera2025PRL} used the measured solid-state clock frequency in \thcaf{}~\cite{Zhang2024-Th229Comb}, together with calculated isomer shifts, to predict the corresponding transition frequencies for trapped-ion platforms.
Their analysis predicts hyperfine-free $\omega_{\rm clk}(\ce{^{229}Th^{3+}})=2\,020\,407\,114(70)~\mathrm{MHz}$ and $\omega_{\rm clk}(\ce{^{229}Th^{4+}})=2\,020\,407\,648(70)~\mathrm{MHz}$.
Thus the ion-clock  searches should not simply scan at the measured solid-state resonance frequency: the expected \Thppp{} and \Thpppp{} nuclear resonances are displaced by hundreds of megahertz by the platform-dependent isomer shift.

The accuracy advantage of the ion platform comes from the absence of static site disorder, defect states, and large inhomogeneous crystal fields.
Dominant systematics instead resemble those of optical ion clocks.
Compared to traditional ion clocks, the nuclear character suppresses multiple environmental sensitivities, but does not remove the need for clock-state engineering.
The cost of accuracy is clock stability and operational complexity.
A single-ion or few-ion clock has a small number of quantum oscillators, and therefore its short-term stability is limited by quantum projection noise and by the cycle time required for state preparation, VUV interrogation, and readout.
There is significant experimental overhead in the trapping and cooling technology required to prepare and manipulate the ion(s).
This is the central trade-off with solid-state clocks: trapped ions are the natural accuracy platform, while crystals are the natural ensemble-stability and portability platform.

\section{Solid-State Nuclear Clocks}
\label{Sec:SolidState}

The solid-state implementation of a nuclear clock offers wholly unique capabilities derived from the ability to embed nearly Avogadro-scale numbers of interrogable nuclei within a rugged crystal host.
The chemical bonding of the solid-state replaces the trap of ion and optical lattice clocks to hold \thor{}, providing both a dramatic simplification in experimental complexity and extraordinary signal-to-noise ratios.
This section reviews the seminal theoretical and experimental advances in laser spectroscopy of \thor{} in solids.

\subsection{Energy shifts and platform-dependent frequencies}
The central metrological question for a nuclear clock is not simply the value of the bare nuclear energy, but how reproducibly that energy is realized in a chosen material environment.
The diagonal part of the electron--nucleus interaction, see Sec~\ref{Sec:El-Nucl-coupled}, organizes the leading platform-dependent shifts.
For \thor{} it is useful to expand this interaction in electromagnetic multipoles,
\begin{equation}
    H_{\rm int}=H_{E0}+H_{M1}+H_{E2}+\cdots .
    \label{eq:shift-multipole-expansion}
\end{equation}
The scalar electric-monopole term $H_{E0}$ gives the isomer shift: a change in electronic sub-system energy caused by the different charge distributions of the nuclear ground and isomeric states.
The magnetic-dipole and electric-quadrupole terms split magnetic sublevels and couple the nucleus to local magnetic fields and electric-field gradients.
In a fluoride crystal, for example, nearby $^{19}$F nuclear moments contribute magnetic broadening, while the charge-compensating defect complex sets the electric-field gradient.
Magnetic and quadrupolar shifts can often be averaged over Zeeman or hyperfine components, but the scalar isomer shift remains as a genuine offset between clock realizations.

\citet{Rellergert2010} gave the first detailed shift budget for a solid-state \thor{} clock.
\rev{\citet{Kazakov2012-performance-229Th-clock}
subsequently studied nuclear spin bath decoherence and a fluorescence-based clock-stabilization scheme.} \citet{Rellergert2010}  emphasized that the solid-state transition frequency should be shifted from the free-ion value by the local electronic density and crystal fields, but that this need not prevent a useful frequency reference if thorium occupies a well-defined lattice site.
The model assumed an ionic, closed-shell host, so that unpaired electronic spins and ordinary electronic currents are absent and the dominant local terms are the nuclear hyperfine couplings generated by the lattice.

Their order-of-magnitude estimates are still a useful scale for current materials work.
The electric-monopole, or contact, contribution was bounded at roughly $100$~MHz relative to a free ion, with a temperature coefficient of order $10$~kHz/K.
With $0.1$~mK temperature uniformity this becomes a hertz scale contribution to the ensemble linewidth.
For the magnetic-dipole term, \citet{Rellergert2010} noted that one nuclear magneton at a typical nearest
neighbor distance of $2.5$~\AA{} produces a field of about $400$~mG.
Using the then-available magnetic moments of the $5/2^+$ ground state and $3/2^+$ isomer, this gives a few hundred hertz from a single close spin and an estimated $1$--$10$~kHz broadening after summing over neighboring nuclei.
The electric-quadrupole term was identified as the most site-dependent shift: typical crystal electric-field gradients of $10^{16}$--$10^{21}$~V/m$^2$ correspond to shifts ranging from roughly $1$~kHz to $100$~MHz, minimized by high local symmetry and common to all nuclei only if the Th dopants occupy a single site.
Lattice vibrations were expected to be less severe: optical phonons involving motion of Th relative to nearest neighbors are largely frozen out, while the second-order Doppler shift was estimated at about $1$~Hz/K.

The relevant shifts can be large in absolute frequency, but they are clock errors only to the extent that they vary
across the ensemble or drift in time.
\citet{Rellergert2010} therefore separated precision from accuracy.
A dense ensemble could support short-term fractional resolution in the range $3\times10^{-17}$--$10^{-15}$ after one second of photon collection, whereas long-term accuracy was expected to be set mainly by thermometry and the temperature dependence of the monopole and
quadrupole shifts, at the level of a few parts in $10^{-16}$ for the assumptions made in that work.
While these early predictions are now largely verified, the precision afforded by recent experimental work~\cite{Thielking2018, Yamaguchi2024,Zhang2024-Th229Comb, Ooi2026, Hiraki2026Science,Morawetz2026Nature} provide missing details, such as the moments of the excited nuclear state and actual crystal defect geometry.
In what follows, we briefly summarize the current understanding of the terms of Eq.~\eqref{eq:shift-multipole-expansion}.

Quantitative analysis requires an accurate description of the electronic structure of the host material.
In doped crystals, \thor{} can occupy several distinct local geometries~\cite{Dessovic2014}.
One such Th-doped geometry in \caf{} is shown in Fig.~\ref{fig:HE2_defect}(b): \ce{Th} substitutes for \ce{Ca^2+} as \ce{Th^4+}, requiring two interstitial \ce{F^-} ions for charge compensation.
The diffuse electronic clouds of the \ce{F^-} ions overlap with the closed-shell \ce{Th^4+} ion, reducing its effective charge to about $3.5$.
Qualitatively, one may view the \thor{} nucleus as encased in the closed-shell \ce{Th^4+} ion, which is itself immersed in the electronic cloud of the valence band.
This ``matryoshka'' picture~\cite{Perera2025PRL,Elwell2025-ThO2,Morgan2025_internal_conversion} provides a bridge between periodic density-functional calculations, which are common in materials science, and {\em relativistic} atomic many-body calculations.
Relativity is essential because Eq.~\eqref{eq:shift-multipole-expansion} must be evaluated in the nuclear region, where the electrons are strongly relativistic: $\alpha Z\approx 0.66$ for \thor{}.

{$H_{E0}$---}
The calculation by \citet{Perera2025PRL} sharpen the understanding of $H_{E0}$ by computing host-dependent isomer shifts for the actual platforms now used in experiments.
The key point is the one emphasized in Sec.~\ref{Sec:El-Nucl-coupled}: the observed clock transition is a transition of the compound electron-nucleus system.
After appropriate averaging over magnetic and quadrupole substructure, the remaining platform dependence is dominated by the scalar isomer shift induced by the change in nuclear charge radius.

\citet{Perera2025PRL} combined relativistic atomic many-body calculations with periodic density-functional calculations of the host electronic structure and showed that a simple M\"ossbauer-style estimate based only on the electron density at the nucleus misses the dominant contribution from valence-band (VB) electrons.
The reason is that the change in nuclear radius first perturbs the inner-shell electrons, thereby modifying the mean field experienced by the valence electrons.
In their formulation, the total isomer shift is decomposed into the contribution from the \ce{Th^4+} ion and a crystal-specific VB contribution.
The latter can be expressed in terms of integrated projected densities of states, as obtained from materials-science codes, and the isomer shifts of the corresponding states of the \ce{^{229}Th^3+} ion.

For current solid-state hosts, \citet{Perera2025PRL} find VB contributions to the isomer shift of order $-250$~MHz, with a spread of about $60$~MHz over the considered materials and defect sites.
Moreover, a high-precision measurement~\cite{Zhang2024-Th229Comb} in \thcaf{} has been translated, after subtracting the calculated isomer-shift difference, into a prediction for nuclear clock frequencies for \Thppp{}, \ce{Th^4+}, and the bare \thor{} nucleus; see Table~\ref{Tab:Th229m-props} and Sec.~\ref{Sec:Ions}.

{$H_{M1}$---}
In a solid, the clock nucleus experiences not only any externally applied field, but also the dipolar fields of nearby host nuclei and paramagnetic defects.
The corresponding differential Zeeman shift is
\begin{equation}
    \hbar\delta\omega_Z(t)
    =
    -\left[
    \bra{m}\hat\mu\ket{m}
    -
    \bra{g}\hat\mu\ket{g}
    \right]\cdot\mathbf B_{\rm loc}(t),
    \label{eq:solid-state-zeeman-noise}
\end{equation}
with the appropriate matrix elements taken in the crystal-field-split nuclear eigenstates and $\mathbf B_{\rm loc}(t)$ being the local magnetic field felt by the \thor{} nucleus.
In \thcaf{}, the nearby $^{19}$F spins are the unavoidable bath, while in \thf{} and \thlisaf{} additional host nuclei also contribute.
Early solid-state clock estimates already identified this nuclear-spin bath as a likely source of kHz-scale broadening, and later clock-performance studies showed that it can limit Ramsey-type interrogation if left untreated~\cite{Rellergert2010,
Kazakov2012-performance-229Th-clock}.

The useful language is the same as in magnetic-resonance line-shape theory.
If the local field is static during the interrogation time, different thorium sites acquire different fixed values of $\delta\omega_Z$ and the ensemble line is inhomogeneously broadened.
In free induction decay (FID) measurements in nuclear magnetic resonance (NMR) experiments, this dephasing due to a variation of Larmor frequency is captured in the effective transverse relaxation time, $T_2^*$.
Broadening of this type can be removed with a spin echo in NMR and appears removable for nuclear clocks.
If the bath evolves during the excitation, the optical coherence is reduced.
Such evolution occurs from spin-phonon coupling and spin-spin coupling.
In NMR experiments, spin-phonon coupling is a primary contributor to the longitudinal relaxation time $T_1$.
In high band gap fluorides, Debye temperatures are typically above room temperature and therefore spin-phonon coupling is sub-dominant, with $T_1$ measured to be a minute or longer.
Spin-spin coupling, however, persists, assuming no polarizing magnetic field is applied, at all but the lowest temperature.
This coupling, depending on its timescale, can be difficult to mitigate, and in NMR it is captured by the pure dephasing time $T_2$.

Previous treatments of this interaction are incomplete.
\citet{Rellergert2010} provided only a simple estimate based on static, inhomogeneous broadening finding transition linewidths of 1-10~kHz.
Meanwhile, \citet{Kazakov2012-performance-229Th-clock} use a master equation approach to find linewidths of roughly 0.1-1~kHz, however, it is well known in the NMR community that the problem is non-Markovian and not compatible with a Lindbladian approach~\cite{Fine2005}.
A simple estimate can be made by combining the inhomogeneous broadening, $\Gamma_{in}$, with measured \ce{CaF2} spin-spin dephasing time of $T_2\sim20~\mu$s~\cite{Meier2012}, suggesting the transition linewidths will be $\sqrt{(\pi T_2)^{-2} + \Gamma_{in}^2}\lesssim20$~kHz.
This is likely an upper estimate as the quadrupole interaction $H_{E2}$ should suppress spin-spin coupling between \ce{Th-F} relative to \ce{F-F}.
More accurate treatments of the problem are possible using the tools of NMR~\cite{Fine2005, Kuprov2019, Kubo1966}.

There are several potential mitigation strategies for this interaction.
First, one may use spectroscopic averaging over opposite Zeeman components or choose transitions whose first-order magnetic shifts cancel to high accuracy.
This is the solid-state analogue of the hyperfine averaging and stretched-state protocols used in ion-clock proposals of Sec.~\ref{Sec:Ions} but here it must also contend with the distribution of crystal fields over sites.
Second, one may polarize, decouple, or dynamically average the host-spin bath.
Third, one can choose a host in which the nuclei surrounding thorium are spinless or nearly spinless.
This last route motivates recent materials-design proposals, including polyatomic-anion hosts and explicitly spinless crystal environments for \thor{} clocks~\cite{Morgan2025-design-of-Th-materials, morgan2025-spinless}.
Fourth, it may be possible to implement a magic angle spinning protocol~\cite{Anisimov2007}.
Fifth, we expect that by controlling lattice geometry and host quadrupole interaction it is likely possible that the spin-spin coupling can be engineered to at least partially control the magnetic broadening.
Thus, this interaction may lead to critical design criteria for the host lattice.

{$H_{E2}$ ---}
In a solid, the clock nucleus also experiences an electric field gradient due to the crystal fields, leading to the familiar quadrupolar interaction $H_Q = \frac{1}{6} \sum_{ij} Q_{ij} V_{ij}$, where $V_{ij} = \partial^2\Phi/(\partial x_i \partial x_j)$ is the electric field gradient (EFG) tensor expressed via the electrostatic potential $\Phi$ at the nucleus. We defined the EFG sign consistent with quantum chemistry and \thor{} literature conventions~\cite{DerPerKro2026-EFGs}. The potential $\Phi$ is generated by all the electrons and other nuclei in the crystal and depends sensitively on the details of the electronic structure.
The EFG tensor is real, symmetric, and traceless, $V_{xx}+V_{yy}+V_{zz}=0$, due to \rev{the} Poisson equation for a point-like nucleus.
It can be diagonalized, defining a local reference frame that depends on the geometry local to the nucleus. The three axes are labeled by convention so that $|V_{zz}| \geq |V_{xx}|\geq |V_{yy}|$, leading to the representation
\begin{equation}
H_{E2} = \frac{Q V_{zz} }{4I(2I - 1)}  \left[ 3I_z^2 - I(I + 1) + \eta \left( I_x^2 - I_y^2 \right) \right]\,,
\label{Eq:HQ-PAF-canonical}
\end{equation}
where  the asymmetry parameter is defined as $\eta \equiv (V_{xx} - V_{yy})/V_{zz}$ and $Q$ is the spectroscopic value of the quadrupole moment; see Table~\ref{Tab:Th229m-props}.
To distinguish between spins and $Q$-moments of the two nuclear states, we introduce a subscript $\alpha =g,m$
in the formulas below.

\rev{In zero magnetic field, the quadrupole interaction splits the ground state into three pairs of degenerate sublevels and the isomer into two.
These pairs are called Kramers doublets.
The field-gradient magnitude $|V_{zz}|$ sets the scale of the splittings, while $\eta$ measures the departure from axial symmetry and controls the mixing of spin projections.
The expressions below give the energies and compositions of these states.}

This interaction is the largest of the hyperfine interactions for insulator materials and it is therefore natural to use its eigenstates as the basis of choice.
Defining $C_\alpha \equiv  Q_\alpha V_{zz}/(4I_\alpha (2I_\alpha -1))$ and an asymmetry mixing parameter $\mathcal{M} \equiv \sqrt{1 + \eta^2/3}$, the excited state manifold splits into two doubly-degenerate states, known as Kramers doublets, which in the $\ket{I,m_I}$ basis are:
\begingroup
\small
\begin{align}
    &E_e\left(\pm\frac{3}{2}\right) = 3 C_e \mathcal{M}~\textrm{with eigenvectors:}\nonumber\\
    &\ket{e, \pm\frac{3}{2}} = \frac{1}{\sqrt{2\mathcal{M}}}\left(\sqrt{\mathcal{M}+1}\ket{\frac{3}{2},\pm\frac{3}{2}} + \sqrt{\mathcal{M}-1}\ket{\frac{3}{2},\mp\frac{1}{2}}\right)\nonumber\\
    &E_e\left(\pm\frac{1}{2}\right) = -3 C_e \mathcal{M}~\textrm{with eigenvectors:}\nonumber\\
    &\ket{e, \pm\frac{1}{2}} = \frac{1}{\sqrt{2\mathcal{M}}}\left(\sqrt{\mathcal{M}+1}\ket{\frac{3}{2},\pm\frac{1}{2}}-\sqrt{\mathcal{M}-1}\ket{\frac{3}{2},\mp\frac{3}{2}}\right)
\end{align}
\endgroup
When $\eta\neq0$, the $\left(\hat{I}_{\alpha,x}^2 - \hat{I}_{\alpha,y}^2\right)$ term of the Hamiltonian mixes states with $\Delta m=\pm 2$, however, since $1\leq \mathcal{M} \leq 2/\sqrt{3}$, these states have a predominant $\ket{I,m_I}$ component and it is customary to label the eigenstate by that value of $m_I$.
When $\eta = 0$, the Hamiltonian is diagonal in the $\ket{I,m_I}$ basis.

For the nuclear ground state, the interaction splits the six $m_I$ components into three Kramer's doublets ($n = 0,1,2$) with energy
\begin{align}
    E_g(n) &= 4\sqrt{7} C_g \mathcal{M} \cos{\left(\frac{\theta -2 n \pi}{3}\right)}\,,\nonumber\\
    \cos\theta &= \frac{10(4-3\mathcal{M}^2)}{7\sqrt{7}\mathcal{M}^3}\,,
\end{align}
with degenerate eigenvectors,
\begin{align}
    \ket{\pm,n} &= \frac{1}{\mathcal{N}(n,\eta)}\left(\sqrt{10}\eta\left(\frac{E_g(n)}{C_g}+2\right)\ket{\frac{5}{2},\pm\frac{5}{2}} \right.\nonumber\\
    &+ \left(\frac{E_g(n)}{C_g}-10\right)\left(\frac{E_g(n)}{C_g}+2\right)\ket{\frac{5}{2},\pm\frac{1}{2}} \nonumber\\
&+\left. 3\sqrt{2}\eta \left(\frac{E_g(n)}{C_g}-10\right)
\ket{\frac{5}{2},\mp\frac{3}{2}}\right)\,,
\end{align}
where $\mathcal{N}(n,\eta)$ is a normalization factor.
The structure can be understood by noting that for $\eta = 0$ the state energies become $\{10 C_g, -2C_g, -8C_g\}$ for $n = \{0,1,2\}$, respectively, leading to unmixed eigenstates in the $\ket{I,m_I}$ basis, i.e. the eigenstates are $\{ \ket{5/2,\pm5/2}, \ket{5/2,\pm3/2},\ket{5/2,\pm1/2} \}$ for $n = \{0,1,2\}$, respectively.
As for the $I = 3/2$ case, when $\eta \neq 0$ the states are predominantly composed of a single $\ket{I,m_I}$ state, which is typically used to label the state.
However, because $m_I$ is no longer a conserved quantity, the M1 selection rules for the isomeric transition are modified by the state mixing.

With this understanding the major features of hyperfine-resolved nuclear spectroscopy can be explained, as shown in Fig.~\ref{fig:HE2_defect}(a).
For a given $V_{zz}$ and $\eta$, the ground and excited states split into three and two, respectively, doubly degenerate states that we label by their primary $\ket{I,m_I}$ component. For typical values, the splitting between Kramer's doublets is $\mathcal{O}(100)$~MHz.
This leads to five transitions between states with $|\Delta m_I|\leq1$, and if $\eta\neq0$ a sixth weak line between $\ket{3/2,\pm1/2}\leftarrow\ket{5/2,\pm5/2}$ states.

\rev{Whether \thor{} is a constituent of a stoichiometric crystal
such as \thf{} or a non-native dopant, the values of $V_{zz}$ and
$\eta$ are set by the local crystal field, which is determined
by the atomic arrangement and the electronic configuration
of \thor{} in the crystal.}
However, in the latter case, since \thor{} is a defect to the crystal, more than one arrangement is often possible~\cite{Jackson2009}.
\rev{As shown in Fig.~\ref{fig:HE2_defect}(b) for \thor{}:\ce{CaF2}, one such defect structure has \thor{} replace a Ca atom and two interstitial \ce{F} atoms appear between typical \ce{F} locations to compensate the charge difference between \thor{}$^{4+}$ and \ce{Ca^2+}.}
Thus, the nuclear spectroscopy becomes a sensitive probe of the crystal structure.
\citet{Hiraki2026Science} used the measured spectrum to conclude that in \thor{}:\ce{CaF2} there were likely four distinct charge compensating arrangements.  Although measured EFGs provide a means of fingerprinting defect structures, systematic benchmarks of EFG accuracy in materials-science codes remain scarce. At low \thor{} density, these crystal defects are non-interacting, however, as the \thor{} density is increased it can be expected that additional lattice strain will occur and given the stochastic nature of the defect arrangement,
the nuclear transitions will be broadened and nuclear clock performance degraded.
Such effects have likely been observed in \citet{Higgins2025-Temperature-dependence-ThCaF2}.

\subsection{Relaxation mechanisms}
In high band gap hosts, $E_{\text{gap}} > \hbar \omega_\mathrm{nuc}$, the isomer decays radiatively.
The ultimate promise of these platforms is to offer the same narrow linewidth as in the ion trap, while offering unmatched clock stability by interrogating $\gtrsim 10^{16}$ nuclei.
Several effects complicate this picture.
The first is a slight increase in the homogeneous linewidth due to an increased density of photonic states inside the crystal.
The lifetime of the radiative decay, $\tau'$, is modified by the index of refraction $n$ of the surrounding medium when compared to the vacuum lifetime $\tau_{\text{vac}}$.
For a predominantly $M1$ decay, as in the case of the \thor{} isomer, in an isotropic material this relation is given by $\tau_{\text{vac}} = n^3 \tau'$~\cite{Nienhuis1976a, Tkalya2001_purcell}.
Given that indices of refraction are of order unity, this is not a significant impediment to the solid-state clock.

However, depending on the local photonic and electronic environment, other pathways, which we refer to as \textit{quenching}, can depopulate, both radiatively and non-radiatively, the isomer more quickly than the unmodified radiative decay.
These pathways must be managed so that clock performance is not degraded.
The observed decay rate in a material can be characterized as,
\begin{equation}
    \Gamma_{\rm obs}
    =
    \Gamma_{\gamma}
    +\Gamma_{\rm IC}
    +\Gamma_{\rm EB}
    +\Gamma_{\rm pq}
    +\Gamma_{\rm xq}
    +\cdots ,
    \label{eq:quenching-rate-budget}
\end{equation}
where the additional terms represent internal conversion, electronic-bridge decay, photo-induced quenching, x-ray induced quenching, and other material relaxation channels.
Whether or not a term is a useful tool depends on the clock platform and interrogation protocol.
For photon-based spectroscopy, quenching reduces the radiative branching ratio and can hide the nuclear signal.
In absorption spectroscopy, if it does not broaden the line more than the other broadening mechanisms it is likely beneficial.
Likewise, for a clock with an electron readout, or for rapid state preparation between interrogation cycles, the same electron-nucleus coupling can be an asset.

The elementary IC process is an off-diagonal counterpart of the
shift physics discussed above, see Sec.~\ref{Sec:El-Nucl-coupled}.
The nucleus makes the transition
$\ket{m}\rightarrow\ket{g}$ while an electron is resonantly promoted to an energetically
allowed final state,
\begin{equation}
    \Gamma_{\rm IC}
    =
    \frac{2\pi}{\hbar}
    \sum_f
    \left|
    \bra{g;f}H_{\rm int}\ket{m;i}
    \right|^2
    \delta\!\left(E_m-E_g+E_i-E_f\right).
    \label{eq:solid-state-ic-golden-rule}
\end{equation}
because all remaining electrons are bound too deeply.
In an ideal wide-band-gap
insulator the IC channel is  closed if no valence, defect, or
conduction-band final state satisfies the energy-conservation condition in
Eq.~(\ref{eq:solid-state-ic-golden-rule}).  This is the simple argument behind
the use of \Thppp{} and fluoride crystals as long-lived radiative platforms.

Real crystals, however, are not described only by their bulk band gap.  Thorium
doping produces charge-compensating defects and local Th-centered electronic
levels.  If an empty defect level lies below the nuclear energy, the isomer can
de-excite by promoting an electron from the valence band into that level; if a
defect level is already occupied, the isomer can promote that electron into a
higher defect or conduction-band state.  The relevant coupling is local and is
therefore strongest for states with significant electronic amplitude at the
thorium site.  This is why the same microscopic defect complex that determines
the isomer shift and electric-field gradient also determines the nonradiative
decay rate.  Early estimates already emphasized the sensitivity of the isomer
lifetime to the electronic environment~\cite{Karpeshin2007}; recent work has
developed the corresponding theory for solid-state hosts and defect-mediated
internal conversion~\cite{Elwell2024-lisaf,Morgan2025_internal_conversion}.

The \thlisaf{} experiments provide a useful example of how this physics appears
spectroscopically.  In addition to the narrow nuclear line, the experiment
observed a broad feature near the isomer energy that produced seconds-scale
fluorescence.  Once the nuclear transition was identified, it became clear that a
large material response near 148~nm does not by itself imply coherent nuclear
excitation.  It can instead reveal an electronic defect manifold that is close in
energy to the nuclear transition.  The same defect manifold can quench the
isomer: nuclear de-excitation creates an electron-hole or defect-hole excitation,
and the radiative photon is not emitted.  This interpretation explains why not
all thorium nuclei need participate in the long-lived radiative signal and why
sample preparation, charge compensation, and radiation damage directly enter the
clock problem~\cite{Elwell2024-lisaf,Morgan2025_internal_conversion}.

Photo-induced quenching is the controlled version of the same idea.  If an
off-resonant laser prepares or admixes an electronic defect excitation, it can
open an otherwise weak internal-conversion pathway.  In \thlisaf{}, illumination
detuned from the nuclear resonance was observed to accelerate isomer decay, and a
model in which photoexcited gap states open an internal-conversion channel
reproduces the measured quenching cross section~\cite{Terhune2025-photo-induced}.
Laser-induced quenching has also been demonstrated in \thcaf{}, where light over
a broad range of wavelengths reduces the effective isomer lifetime
~\cite{Schaden2024-Th-quenching}.  These observations recast quenching from a
mere loss mechanism into an active control knob: it can shorten the clock cycle,
enable state reset, and provide a diagnostic of otherwise hidden defect states.

The material details matter.  In \thcaf{}, thorium radioactivity and VUV
illumination can create color centers and alter the fluorine stoichiometry,
which increases absorption near the clock wavelength and motivates careful
growth, annealing, fluorination, and temperature control~\cite{Beeks_2024}.
Cooling can help suppress damage and stabilize the optical response, but it also
adds cryogenic complexity.  By contrast, the \thlisaf{} photoquenching
experiments show that useful nuclear excitation and controlled depopulation can
be obtained without the same degree of low-temperature mitigation
~\cite{Elwell2024-lisaf,Terhune2025-photo-induced}.  The comparison illustrates why
host selection is not only a question of band gap: the relevant figure of merit
is the full defect and damage landscape at the thorium site.

Finally, quenching can be deliberately exploited for readout.  In \thorox{}, the
material is opaque to 148~nm light over macroscopic distances, but nuclei close
to the surface can still be excited and then relax by internal conversion into
electron-hole excitations.  The ejected electrons can be detected,
leading to a fast, solid-state readout channel rather than photon counting
~\cite{Elwell2025-ThO2,Morgan2025_internal_conversion}.  This possibility changes the clock
design trade-off.  A radiative-decay fluoride clock aims to preserve the long nuclear
lifetime and narrow linewidth; an internal-conversion readout scheme sacrifices
that lifetime after interrogation in exchange for rapid reset and a potentially
large electron emission current.

\section{Exotic physics probes}
\thor{} is expected to be exceptionally sensitive to variations of fundamental constants of nature (FCs), such as particle masses or the fine-structure constant $\alpha$.
The Standard Model of elementary particles fixes the values of these constants, but its extensions may promote them to dynamical fields.
Atomic clocks, whose transition frequencies depend on FCs, have already proven to be sensitive probes of such variations~\cite{SafBudDeM2018.RMP}.

Clock sensitivity to FC variations is conventionally parameterized in terms of sensitivity coefficients $K_X$,
\begin{equation}
\frac{\delta\omega_{\rm clk}}{\omega_{\rm clk}}
=
\sum_X K_X \frac{\delta X}{X},
\end{equation}
where $X$ runs over the constants considered.
For conventional atomic clocks, $K_\alpha \sim \mathcal{O}(1)$, but it is expected to be orders
of magnitude larger for \thor{}.
The isomer energy is only
$\sim 8.4$~eV, arising from a near cancellation of MeV-scale
electromagnetic and strong-interaction contributions,
\begin{equation}
\hbar\omega_{\rm nuc}
=
\Delta E_{\rm nuc}+\Delta E_{\rm EM}
\ll
|\Delta E_{\rm nuc}|,|\Delta E_{\rm EM}|.
\end{equation}
A small variation in either term shifts this delicate balance and is therefore amplified in the measured value of $\omega_{\rm nuc}$.
Qualitatively, the fractional sensitivity is $K_{\rm canc}\sim |\Delta E_{\rm nuc,EM}|/(\hbar\omega_{\rm nuc})$, which can be as large as $10^4$--$10^5$ for nuclear-scale energies.
This simple estimate is the origin of the claim that the \thor{} transition can be orders of magnitude more sensitive to variations of $\alpha$ or of dimensionless strong-sector parameters than atomic clock transitions~\cite{Flambaum2006a}. For most atomic clocks, the dependence on nuclear parameters enters only indirectly, through reduced-mass, hyperfine, or field-shift effects.
The \thor{} transition instead probes nuclear binding directly, which is why it is especially attractive.

The numerical values of the sensitivity factors are, however, a nuclear-structure problem under active investigation.
\citet{FadBerFal2020-ThClockAlphaVar} used charge-radius information together with a constant nuclear density ansatz to infer $K_{\alpha}=-(0.82\pm0.25)\times10^4$.
Recent VUV-comb spectroscopy in \thcaf{} resolved the quadrupole structure of the nuclear transition, providing a more direct handle on the change of nuclear shape.
Using these data in a prolate-spheroid model, \citet{Beeks2024} found $K_{\alpha}=5900(2300)$, with the uncertainty dominated by the
mean-square charge-radius difference and with possible octupole contributions still to be constrained.
A complementary analysis~\cite{Caputo2024a} found likely $\sim 10^4$  enhancements in both geometric and halo-inspired models, while emphasizing that poorly known higher multipoles and skin-thickness changes can in principle reduce the enhancement in finely tuned ``nightmare'' scenarios.
Thus the robust conclusion is not a single exact value of $K$, but a strong expectation of a large enhancement.

This perspective also explains why pre-clock spectroscopy can already be useful for new-physics searches.
A fractional line-position sensitivity that is far from the ultimate clock goal can still correspond to a much smaller effective variation of nuclear parameters once  the $10^4$ sensitivity factors are included.
Recent analyses of \thor{} spectroscopy have used this idea to search for ultralight dark matter through either slow time-dependent line shifts or changes in the observed line shape~\cite{Fuchs2025,Arakawa2026}.

As nuclear-clock technology matures, trapped-ion and solid-state platforms are likely to become complementary probes of FC variation.
Trapped ions offer maximal control of systematic shifts and are therefore natural for precision frequency comparisons searching for slow drifts or coherent oscillations.
Solid-state clocks, by contrast, trade some environmental control for enormous numbers of nuclei and potentially high measurement bandwidth, making them attractive for fast-transient searches.
Examples include short-duration FC excursions induced by ``clumpy'' dark matter~\cite{DerPos14} or by exotic low-mass fields plausibly  emitted in energetic cosmic events such as black-hole mergers~\cite{dailey2020ELF.Concept}.

\begin{acknowledgments}
We are grateful to the participants of the Kavli Institute for Theoretical Physics (KITP) Program “Thorium-229 Nuclear Clocks” for many insightful discussions that helped shape this review.
\rev{We thank Benjamin Katz for helpful corrections to the manuscript.}
We also gratefully acknowledge NSF Grant No. PHY-2309135 to KITP.
This work was supported in part by NSF awards PHYS-2013011 and PHY-2513134, ARO award W911NF-11-1-0369.
ERH acknowledges institutional support by the NSF QLCI Award OMA-2016245.
\end{acknowledgments}

%

\end{document}